\documentclass[acmsmall]{acmart}
\usepackage{multirow}
\usepackage{CJKutf8}
\usepackage{booktabs}
\usepackage{graphicx}
\usepackage{amsmath}
\usepackage{algorithmic}
\usepackage[ruled,linesnumbered]{algorithm2e}
\usepackage{amsfonts}
\usepackage[justification=centering]{caption}
\AtBeginDocument{%
  \providecommand\BibTeX{{%
    \normalfont B\kern-0.5em{\scshape i\kern-0.25em b}\kern-0.8em\TeX}}}

\setcopyright{acmlicensed}
\copyrightyear{xxx}
\acmYear{xxx}
\acmDOI{XXXXXXX.XXXXXXX}

\acmJournal{TOIS}
\acmVolume{0}
\acmNumber{0}
\acmArticle{0}
\acmMonth{0}

\begin{document}

\title{CETN: Contrast-enhanced Through Network for CTR Prediction}

\thanks{This work was supported by the National Natural Science Foundation of China [No.62272001, No.62206002]. The corresponding author of this paper is Yiwen Zhang.}

\author{Honghao Li}
\email{salmon1802li@gmail.com}
\affiliation{%
  \institution{Anhui University}
  \streetaddress{111 Jiulong Rd}
  \city{Hefei}
  \state{Anhui Province}
  \country{China}
  \postcode{230601}
}

\author{Lei Sang}
\affiliation{%
  \institution{Anhui University}
  \streetaddress{111 Jiulong Rd}
  \city{Hefei}
  \state{Anhui Province}
  \country{China}
  \postcode{230601}
}
\email{sanglei@ahu.edu.cn}

\author{Yi Zhang}
\affiliation{%
  \institution{Anhui University}
  \streetaddress{111 Jiulong Rd}
  \city{Hefei}
  \state{Anhui Province}
  \country{China}
  \postcode{230601}
}
\email{zhangyi.ahu@gmail.com}

\author{Xuyun Zhang}
\affiliation{%
 \institution{Macquarie University}
 \streetaddress{Balaclava Rd}
 \state{Macquarie Park NSW}
 \country{Australia}
 \postcode{2109}
}
\email{xuyun.zhang@mq.edu.au}

\author{Yiwen Zhang} 
\affiliation{%
  \institution{Anhui University}
  \streetaddress{111 Jiulong Rd}
  \city{Hefei}
  \state{Anhui Province}
  \country{China}
  \postcode{230601}
}
\email{zhangyiwen@ahu.edu.cn}

\renewcommand{\shortauthors}{H. Li et al.}

\begin{abstract}
Click-through rate (CTR) Prediction is a crucial task in personalized information retrievals, such as industrial recommender systems, online advertising, and web search. Most existing CTR Prediction models utilize explicit feature interactions to overcome the performance bottleneck of implicit feature interactions. Hence, deep CTR models based on parallel structures (e.g., DCN, FinalMLP, xDeepFM) have been proposed to obtain joint information from different semantic spaces. However, these parallel subcomponents lack effective supervision and communication signals, making it challenging to efficiently capture valuable multi-views feature interaction information in different semantic spaces. To address this issue, we propose a simple yet effective novel CTR model: Contrast-enhanced Through Network for CTR (CETN), so as to balance the diversity and homogeneity of feature interaction information. Specifically, CETN employs product-based feature interactions and the augmentation (perturbation) concept from contrastive learning to segment different semantic spaces, each with distinct activation functions. This improves diversity in the feature interaction information captured by the model. Additionally, we introduce self-supervised signals and through connection within each semantic space to ensure the homogeneity of the captured feature interaction information. The experiments and research conducted on four real datasets demonstrate that our model consistently outperforms twenty baseline models in terms of AUC and Logloss.
\end{abstract}



\received{XXXXXXXXX}
\received[revised]{XXXXXXXXX}
\received[accepted]{XXXXXXXXXXX}

\maketitle

\section{Introduction}
\label{Section 1}
Accurately predicting user responses to items (e.g., products, movies, and advertisements) under certain contexts (e.g., websites, and apps) plays a pivotal role in commercial personalized information retrieval (IR) scenarios  \cite{widedeep, DIN, DIEN, finalmlp, dcnv2}. The user's probability of clicking on an item serves as an intuitive evaluation metric, widely applied across various downstream scenarios, such as online advertising  \cite{ad1, ad2, deeplight}, recommender systems  \cite{sgl,lightgcn, FM}, and web search  \cite{ws1, ws2}. Its primary objective is to assess the probability of users clicking on recommended items within a system. On the one hand, the revenue of the vast majority of commercial IR systems is closely tied to user click-through rates (CTR), making the accuracy of CTR prediction directly impact the profitability of these systems. On the other hand, user satisfaction is closely linked to the performance metrics of recommender system. A well-performing CTR prediction model aids in swiftly discerning user interests, thereby enhancing the user experience.

Capturing effective feature interactions is one of the crucial strategies to enhance the performance of CTR models. In the initial period, researchers employ explicit product-based feature interaction to build models  \cite{FM, FFM, pnn1, pnn2}, aiming to mitigate data sparsity issues. However, due to model complexity constraints, they could only capture low-order feature interactions in practical applications  \cite{dcn, fignn, FM, FwFM}. With the recent proliferation of deep learning, deep learning-based CTR models have emerged, where the multi-layer perceptron (MLP) is widely employed  \cite{dcn,dcnv2,deepfm,finalmlp,pnn1,masknet}. MLP implicitly captures high-order feature interaction information. Nevertheless, some existing work has pointed out that the expressive power of a single MLP is inefficient for capturing feature interaction information, and it may even struggle to learn simple inner-product operations  \cite{mlpeffective,neuralvsmf}. To overcome the performance bottleneck of MLPs in capturing feature interaction information, researchers have attempted to combine explicit product-based feature interactions with the ability of MLPs to implicitly capture high-order feature interactions, leading to significant performance improvements  \cite{openbenchmark}. Depending on the integration approach, this can be divided into two structures, namely the parallel structure and the stacked structure \cite{EDCN, finalmlp, deepfm, pnn1}. The stacked structure attempts to first perform explicit product-based feature interactions on the features and then feed them as input to a MLP \cite{pnn1, masknet, Xcrossnet, dcnv2}. In contrast, the parallel structure jointly trains explicit and implicit components of feature interactions in a parallel manner, utilizing a fusion layer to obtain joint information from different semantic spaces  \cite{finalmlp,deepfm,xdeepfm,dcnv2,masknet}. However, many existing CTR models suffer from three main issues, even if they use the two structures mentioned above, still result in a narrow semantic space available for capturing information, and are unable to efficiently capture diversiform feature interaction information in different semantic spaces, as detailed follow.

\begin{figure*}[t]
  \centering
  \includegraphics[width=1.0\textwidth]{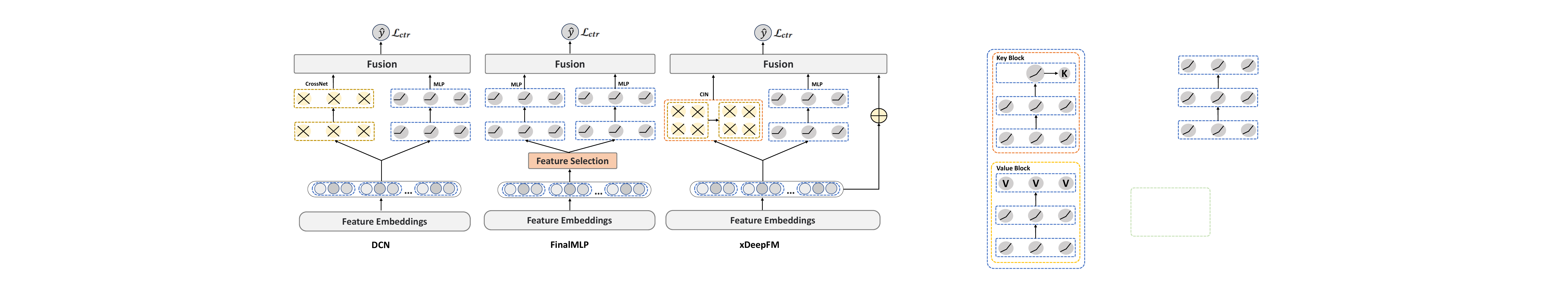}
  \captionsetup{justification=raggedright}
  \caption{\textbf{The architecture of three strong baseline models with parallel structure: DCN, FinalMLP, and xDeepFM.}}
  \label{fig1}
\end{figure*}

\textbf{Lackluster segmentation in the semantic space.} Most CTR models with parallel structures share embedding layers among their parallel subcomponents and directly use concatenated embeddings as inputs for the model  \cite{dcn,dcnv2,deepfm,xdeepfm}. This results in the information from different semantic spaces remaining similar, and relying solely on the information capture capacity of a single subcomponent in an attempt to improve diversity in the obtained information is undoubtedly inefficient. For models that employ parallel subcomponents with the same structure  \cite{CL4CTR,finalmlp,masknet}, the problem of semantic space segmentation becomes even more pronounced. Without capturing feature interaction information in different semantic spaces differently, it often leads to sub-optimal performance and fails to guarantee the diversity of feature interaction information.

\textbf{Feeble supervisory signals in multi-semantic space.} CTR models based on parallel structures often require a fusion layer to aggregate the feature interaction information captured by various subcomponents to obtain the final prediction  \cite{finalmlp,masknet,dcn,xdeepfm}. However, prior to information aggregation, there is a lack of communication and effective supervisory signals between the individual subcomponents. This hinders the model from capturing high-quality, non-redundant information and diminishes the diversity of the information captured.

\textbf{Excessive noise in a multi-semantic space.}
While we pursue diversity within semantic spaces, it inevitably leads to the challenge of "being too different". Empirically, we have observed that if different parallel subcomponents exclusively capture entirely distinct information, it results in the model aggregating a substantial amount of noise signals in the fusion layer, thereby diminishing model performance. Hence, ensuring homogeneity in the information captured by individual subcomponents is crucial.

To explain the three challenges we proposed in more detail, as illustrated in Figure \ref{fig1}, we present three strong baseline models as examples. In the case of DCN  \cite{dcn}, its two subcomponents share an embedding layer. It explicitly captures low-order feature interactions using CrossNet and implicitly captures high-order feature interactions through an MLP. In the fusion layer, a simple summation is used to aggregate information from two semantic spaces. However, this straightforward information capture approach fails to address the three challenges we have identified. FinalMLP  \cite{finalmlp} begins by segmenting two semantic spaces using a feature selection layer, ensuring diversity in the information contained within these spaces. It then employs two MLPs to implicitly capture feature interaction information within each semantic space. Finally, an explicit product operation is introduced in the fusion layer. This model resolves the issue of semantic space segmentation, ensuring information diversity in these spaces and achieving outstanding performance. However, it lacks effective supervisory signals within each semantic space, making it difficult to ensure that MLP can capture diverse information in semantic space through a self-supervised manner. It also lacks differentiation between primary and auxiliary semantic spaces, failing to guarantee the homogeneity of captured information across the model. Similarly, xDeepFM  \cite{xdeepfm} divides data into three semantic spaces and models feature interactions in a more complex manner to ensure diversity in the captured feature interaction information. However, it faces challenges similar to DCN, which cannot guarantee the diversity and homogeneity of the captured information.

To address the widespread issues among parallel subcomponents, we introduce a novel CTR prediction model in this paper, named \textbf{Contrast-enhanced Through Network (CETN)}. Specifically, to improve diversity in the captured feature interaction information, we employ the semantic space using the product \& perturbation paradigm, then utilize multiple Key-Value Blocks with different activation functions as parallel subcomponents of the model. Furthermore, to address issues stemming from diversity, we propose a Through Network to ensure homogeneity in the information captured by individual subcomponents. Before the fusion layer, we introduce Denominator-only InfoNCE (Do-InfoNCE) and cosine loss to self-supervised learning processes within each semantic space. Extensive experiments conducted on four real-world datasets demonstrate that this simple yet efficient model consistently outperforms twenty baseline models in terms of both AUC and Logloss.

In summary, the primary contributions of this work can be summarized as follows:
\begin{itemize}
\item By analyzing the existing CTR models with parallel structures, we summarize three challenges common to these models, i.e., lackluster segmentation, feeble supervisory signals, and excessive noise. To address these three challenges, we introduce the complementary principles of diversity and homogeneity, leading to the proposal of a new model, CETN.
\item To enhance the diversity of information captured by the model, we further segment the semantic space, utilize Key-Value Blocks with different activation functions as subcomponents, and introduce Do-InfoNCE to provide auxiliary signals for capturing feature interactions.
\item To ensure the homogeneity of the information captured by the subcomponents, we introduced element-wise Through Connection and cosine loss as communication bridges between multiple semantic spaces.
\item We conduct fair and extensive experiments on four real public benchmark datasets to evaluate the performance of our proposed model. Based on the experimental results, we demonstrate that CETN consistently outperforms twenty state-of-the-art baseline models.
\end{itemize}

\section{PRELIMINARIES}
\subsection{Problem Definition}
\label{section 3.1}
The click-through rate prediction task aims to enhance the profitability and user satisfaction of commercial recommendation systems by predicting the likelihood of a current user clicking on items recommended by the system. Therefore, about the CTR prediction tasks based on feature interaction, we can define it formally as follows.

\textit{DEFINITION 1 (CTR Prediction Based on Feature Interactions).} For a given user $u$ and item $v$, three groups of features are extracted from them:
\begin{itemize}
\item  \emph{User profiles} ($x_p$): age, gender, occupation, etc.
\item \emph{Item attributes} ($x_a$): brand, price, category, etc.
\item \emph{Context} ($x_c$): timestamp, device, position, etc.
\end{itemize}

As shown in Table \ref{table 1}, in general, $x_p$, $x_a$, and $x_c$ are multi-field categorical data and are represented using one-hot encoding, and then we utilize an embedding layer to transform them into low-dimensional dense vectors: $\mathbf{e}_i=\textit{E}_i x_i$, where $\textit{E}_i \in \mathbb{R}^{d \times s_i}$ and $s_i$ separately indicate the embedding matrix and the vocabulary size for the $i$-th field, $d$ represents the embedding dimension. Analogously, for numerical data (e.g., age, price), we typically start by discretizing them into categorical data using bucketing method\footnote{\url{https://www.csie.ntu.edu.tw/~r01922136/kaggle-2014-criteo.pdf}} \cite{openbenchmark, GDCN} and then transform them into embedding vectors using embedding methods for categorical features. Subsequently, we can obtain the result representation of the embedding layer: $\mathbf{E}=\left[\mathbf{e}_1, \mathbf{e}_2, \cdots, \mathbf{e}_f\right]$, where $f$ denotes the number of fields.

\begin{table*}[htbp]
\centering
\caption{An example of multi-field categorical data.}
\label{table 1}
\resizebox{\linewidth}{!}{
\begin{tabular}{c|ccc|ccc|ccc}
\hline
\multirow{2}{*}{Click} & \multicolumn{3}{c|}{User ($x_p$)} & \multicolumn{3}{c|}{Item ($x_a$)}  & \multicolumn{3}{c}{Context ($x_c$)}     \\ \cline{2-10} 
                       & age & gender & occupation & brand   & price & category & timestamp & device  & position  \\ \hline
1                      & 30  & female & engineer   & Armani  & 200   & clothing & 2022/8/19  & Huawei  & London    \\
0                      & 20  & male   & student    & Vuitton & 1000  & clothing & 2022/11/3 & Apple   & New York  \\
1                      & 40  & male   & engineer   & Gucci   & 600   & clothing & 2022/3/16 & Samsung & Hong Kong \\
0                      & 20  & female & student    & Dior    & 2000  & cosmetic & 2022/5/26 & Lenovo  & Tokyo     \\ \hline
Number                 & 10  & 2      & 500        & 1000    & 5000  & 1000     & 365       & 1000    & 1000      \\ \hline
\end{tabular}}
\end{table*}

Variable $y \in \{0, 1\}$ is an associated label for user click behavior:
\begin{equation}
    y= \begin{cases}1, & u \text { has clicked } v, \\ 0, & \text { otherwise. }\end{cases}
\end{equation}

Let $o^{(1)}$ denotes $\mathbf{E}$ (namely, the first-order feature interaction representation). If we need to obtain higher-order feature interactions we can use the following form:
\begin{equation}
    o^{(i+1)}=g(o^{(1)}, o^{(i)}),
\end{equation}
where $g(\cdot)$ denotes an interaction function and $o^{(i)}$ denotes the $i$-th order feature interactions. Therefore, for a simple CTR model, the common framework is formulated as:
\begin{equation}
    \hat{y}=\text{Model}(u, v, \{o^{(1)}, o^{(2)}, \dots, o^{(n)}\} ; \theta),
\end{equation}
where Model is the CTR model with parameters $\theta$, and $o^n$ represents feature interactions of appropriate order.

\textit{DEFINITION 2 (Semantic Space in CTR).} Consider an enhanced embedding as an additional auxiliary semantic space, and the original embedding as the original semantic space. For any semantic space, we can define it as follows:
\begin{equation}
    space_i = \text{segment}(\mathbf{E}),
\end{equation}
where the segment represents various enhancement or no operations, such as gating mechanisms, masking mechanisms, noise, and so on.

\textit{DEFINITION 3 (Capturing Feature Interactions from Multi-Semantic Spaces).} Learning feature interactions in only one semantic space can result in a limited range of available feature interaction information. Therefore, we need to expand the semantic space further, as shown in Figure \ref{figure 2}. We further refine the basic CTR model, enabling its formalization as follows:
\begin{figure*}[t]
  \centering
  \includegraphics[width=0.4\textwidth]{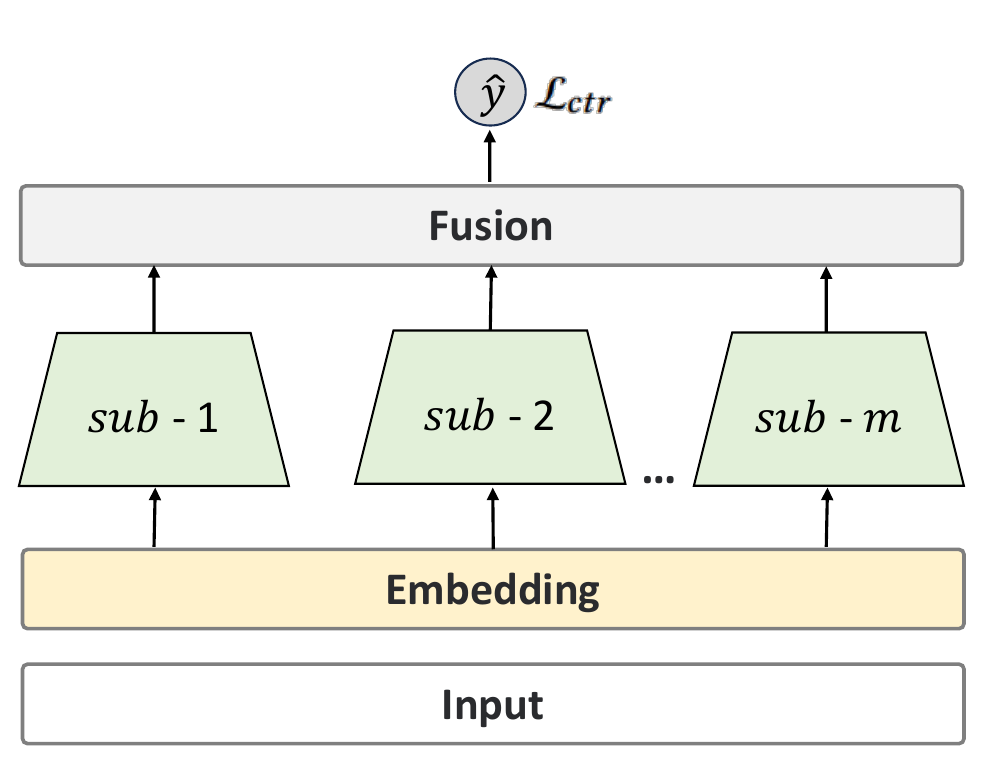}
  \captionsetup{justification=raggedright}
  \caption{The primary backbone structures of common CTR prediction models}
  \label{figure 2}
\end{figure*}
\begin{equation}
    sub_i=\text{subcomponent}_i(u, v, \{o^{(1)}, o^{(2)}, \dots, o^{(s)}\} ; \theta_i), 
\end{equation}
where $sub_i$ denotes the feature interaction information implicit in the $i$-th semantic space, and $o^s, \theta_i$ denote the appropriate feature interaction order and parameters in the current semantic space, respectively. After that, we utilize the fusion layer to further aggregate information from each semantic space, yielding the ultimate prediction.
\begin{equation}
    \hat{y}=\text{fusion}(sub_1, sub_2, \dots, sub_m),
\end{equation}
where $m$ represents the number of suitable semantic spaces, and fusion denotes the aggregate function.

\textit{DEFINITION 4 (Diversity and Homogeneity).} The concepts of diversity and homogeneity have garnered attention in sociology \cite{DH1, DH2}. Sociologists have delved deeply into these two concepts, seeking to comprehend different aspects and dimensions within social structures. Diversity underscores differences among individuals or groups, emphasizing the uniqueness of each member. Conversely, homogeneity emphasizes commonality and similarity, highlighting shared features and commonalities among social members. The dynamic balance between these complementary attributes constitutes diverse and interconnected social systems, aiding sociologists in gaining a better understanding of the complexity of society \cite{DH3, DH4}.

\begin{figure*}[t]
    \begin{minipage}[t]{0.33\linewidth}
        \centering
        \includegraphics[width=\textwidth]{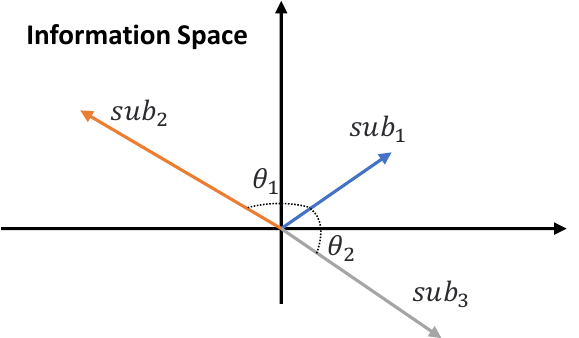}
        \centerline{(a) Excessive diversity}
    \end{minipage}%
    \begin{minipage}[t]{0.33\linewidth}
        \centering
        \includegraphics[width=\textwidth]{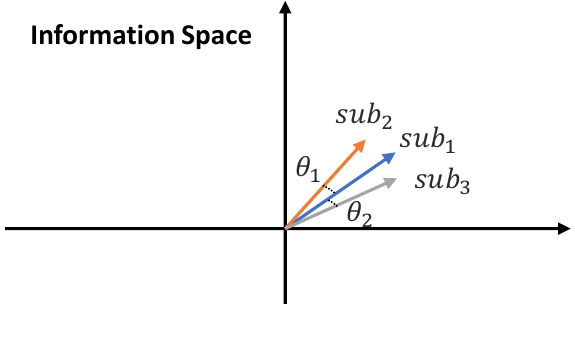}
        \centerline{(b) Excessive homogeneity}
    \end{minipage}%
    \begin{minipage}[t]{0.33\linewidth}
        \centering
        \includegraphics[width=\textwidth]{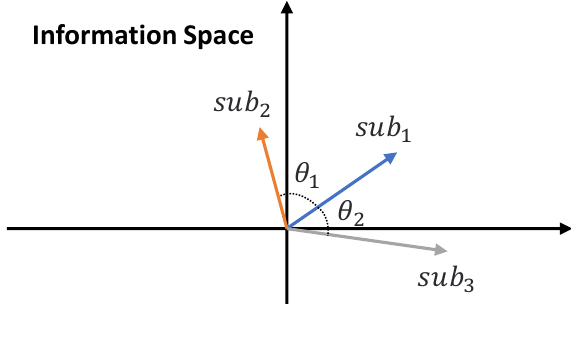}
        \centerline{(c) Achieving a certain balance.}
    \end{minipage}
    \captionsetup{justification=raggedright}
    \caption{An illustration of the diversity and homogeneity in $\mathbb{R}^{2}$.}
    \label{DH}
\end{figure*}

Motivated by the insights from the aforementioned studies, we introduce these two concepts to assist the model in capturing better feature interaction information. Specifically, we can incorporate additional self-supervisory signals in the fusion layer to balance both the diversity and homogeneity of information across various semantic spaces. To illustrate the benefits of this approach more intuitively, we can illustrate these two concepts in $\mathbb{R}^{2}$. In Figure \ref{DH}, if the model captures feature interaction information that is too dissimilar across different semantic spaces, it results in the scenario depicted in Figure \ref{DH} (a). Distinct subspaces acquire significantly different information, leading to excessively large angles between $\theta_1$ and $\theta_2$. Conversely, as shown in Figure \ref{DH} (b), if the model captures overly similar information, redundant information is generated, resulting in suboptimal performance. Figure \ref{DH} (c) represents a balanced scenario. Assuming $sub_1$ is the semantic space with the highest information quality, we can use self-supervisory signals to compel $sub_2$ and $sub_3$ to converge towards $sub_1$ (i.e., reducing the angles $\theta_1$ and $\theta_2$), while preserving their individual information to some extent.

\subsection{Contrastive Learning}
Contrastive learning (CL) is one of the mainstream methods in self-supervised learning  \cite{cl1, cl2, simGCL, XsimGCL}, which first \textbf{augments} the input samples to obtain representations from multiple perspectives, then tries to encourage consistency between pairs of positive samples (\textbf{alignment}) and minimize the agreement between pairs of negative samples (\textbf{uniformity}), ultimately enhancing the model's performance. InfoNCE  \cite{InfoNCE} loss plays a pivotal role in CL, in which the CL loss is:
\begin{equation}
\label{6}
   \mathcal{L}_{c l}=\sum_{i \in \mathcal{B}}-\log \frac{\exp \left(\text{sim}(\mathbf{x}_i^{\prime}, \mathbf{x}_i^{\prime \prime}) / \tau\right)}{\sum_{j \in \mathcal{B}} \exp \left(\text{sim}(\mathbf{x}_i^{\prime}, \mathbf{x}_j^{\prime \prime}) / \tau\right)},
\end{equation}
where $\mathcal{B}$ is the batch size, $x^\prime$, $x^{\prime\prime}$ are input instance representations learned from two different augmentations, $\tau$ represents the temperature coefficient, $\text{sim}(\cdot, \cdot)$ is employed to evaluate the mutual information score between them. However, as pointed out in some previous works  \cite{MISS, simGCL, XsimGCL}, the augmentation and alignment operations on positive samples are not always effective. Therefore, we redefine a contrastive loss more suitable for multi-semantic space learning, which will be presented in Section \ref{section 4.2}.

\begin{figure*}[t]
  \centering
  \includegraphics[width=1.0\textwidth]{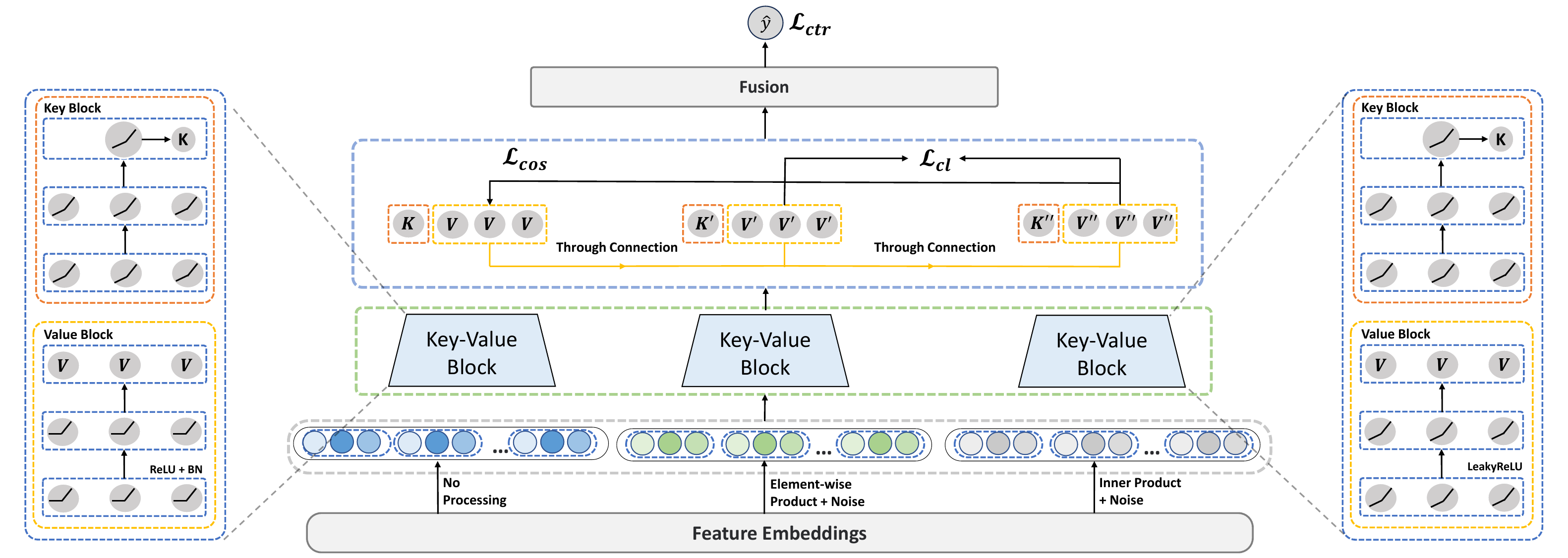}
  \captionsetup{justification=raggedright}
  \caption{The architecture of CETN}
  \label{figure 3}
\end{figure*}

\section{Contrast-enhanced Through Network (CETN)}
\label{Section 4}
 In this section, we will introduce the Contrast-enhanced Through Network (CETN) model. We will approach it from three perspectives and describe the architecture of the CETN model in detail.

\subsection{How to Simply and Efficiently Capture Feature Interactions from Multi-Semantic Spaces}
Most existing parallel-structured CTR models attempt to capture feature interaction information from multi-semantic spaces using \textbf{subcomponents}, such as DCN  \cite{dcn}, DCNv2  \cite{dcnv2}, AutoInt  \cite{autoint}, xDeepFM  \cite{xdeepfm}, and EDCN  \cite{EDCN}. From their model architectures, we can learn that if we want to efficiently capture feature interaction information from multi-semantic spaces, we need multiple subcomponents to model feature interactions in parallel. However, as we pointed out in Section \ref{Section 1}, relying only on the ability of subcomponents to capture information is undoubtedly inefficient. Therefore, in order to better capture the feature interaction information in a multi-semantic space, we need to further \textbf{segment} the information implicit in the semantic space. What can be further considered is that we also need a suitable fusion layer to aggregate information from multi-semantic spaces. Some existing work performs simple summing  \cite{deepfm,deeplight,widedeep} or more complex operations  \cite{finalmlp,fignn,EDCN} (concat, product, linear transform) as a fusion layer for the outputs of each subcomponent, but does not take into account the information weights of the semantic spaces. Therefore, we need a spatial-level attention mechanism to learn the appropriate \textbf{weights} for each semantic space.

In summary, we identify three key elements for extracting feature interaction information from multiple semantic spaces: subcomponent, segmentation, and weight. So, how do we construct a simple and effective CTR model from these elements?

Firstly, considering the efficiency and flexibility of MLP, which performs well in parallel, we utilize it as the subcomponent of the model in each semantic space, and capture information in each semantic space:
\begin{equation}
\begin{aligned}
&\mathbf{v}\ = \textbf{MLP}_{v}\left(\mathbf{E}\right), \\
&\mathbf{v^\prime} = \textbf{MLP}_{v}^\prime\left(\mathbf{S}_{EP}\right), \\
&\mathbf{v^{\prime\prime}} = \textbf{MLP}_{v}^{\prime\prime}\left(\mathbf{S}_{IP}\right),
\end{aligned}
\end{equation}
where $\textbf{MLP}_{v}$ mentioned here is a customizable multi-layer perceptron that uses LeakyReLU as an activation function for its hidden layer and no activation function for its output layer, $\mathbf{S}$ denotes the augmented embeddings, $\mathbf{v}$ is a scalar and represents the final real-values of the feature interaction information in the current semantic space. 

\begin{figure*}[t]
  \centering
  \includegraphics[width=1.0\textwidth]{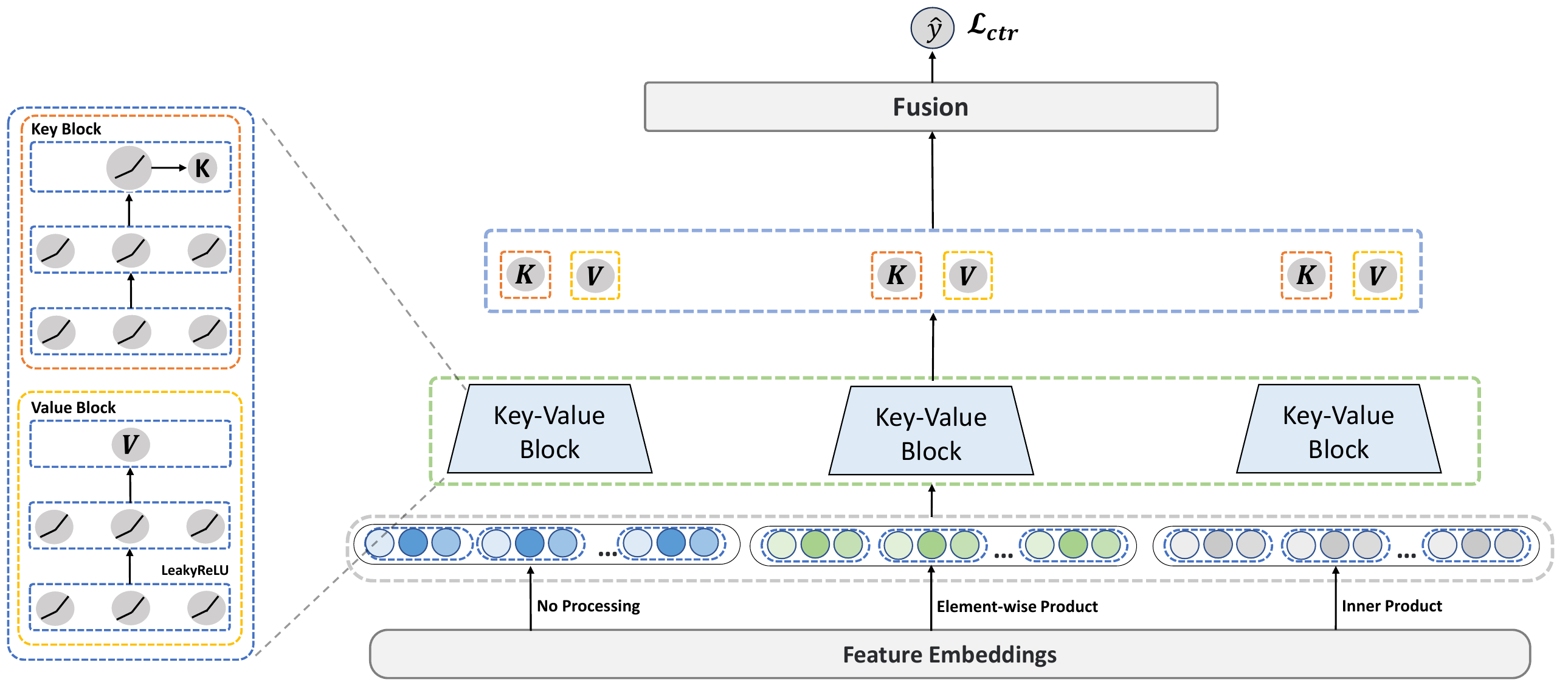}
  \captionsetup{justification=raggedright}
  \caption{The architecture of simMHN}
  \label{figure 4}
\end{figure*}

\begin{table}[t]
\centering
\caption{\textbf{Performance comparison of simMHN with three strong baseline models.}}
\label{table 2}

\begin{tabular}{c|cc|cc|cc}
\hline
\multirow{2}{*}{Models}     & \multicolumn{2}{c|}{Avazu}                                  & \multicolumn{2}{c|}{Criteo}  & \multicolumn{2}{c}{Movielens}         \\ \cline{2-7} 
 & Logloss$\downarrow$   & AUC(\%)$\uparrow$   & Logloss$\downarrow$   & AUC(\%)$\uparrow$          & Logloss$\downarrow$   & AUC(\%)$\uparrow$          \\ \hline
DNN (base)                        & 0.372142                     & 79.2717                      & 0.438233                     & 81.3728      & 0.208941  & 96.9276               \\ 
DCN                         & 0.372353                     & 79.3142                      & 0.438091                     & 81.4103      & 0.204643  & 97.0140               \\
FinalMLP                    & 0.372084                     & \underline{79.3177}                      & \textbf{0.437631}       & \textbf{81.4472}     & \textbf{0.196641}  & \textbf{97.2373}                \\
xDeepFM                     & \underline{0.371944}                     & 79.3121                      & 0.437820                     & 81.4291         & 0.206501  & 96.9769            \\ \hline
simMHN                     & \textbf{0.371091}                     & \textbf{0.794810}                 & \underline{0.437681}                     & \underline{81.4355}     & \underline{0.199238}    &\underline{97.0165}                 \\ \hline
\end{tabular}
\end{table}

Secondly, the product operation is highly effective for modeling feature interactions  \cite{pnn2, aim, Autofeature}. Different product operations often yield distinct feature interaction information. Hence, we conduct pairwise product operations on the feature embeddings, represented as segmented $\mathbf{E}$, to yield $\mathbf{S}$.
\begin{equation}
\begin{aligned}
\mathbf{S}_{EP} &= \Vert_{i=1}^n \Vert_{j=i}^n\ \mathcal{F}_{EP}(\mathbf{e_i} , \mathbf{e_j}), \\
\mathbf{S}_{IP} &= \Vert_{i=1}^n \Vert_{j=i}^n\ \mathcal{F}_{IP}(\mathbf{e_i} , \mathbf{e_j}),
\end{aligned}
\end{equation}
where $\Vert$ denotes the concat operation, $\mathcal{F}_{(\cdot)}$ indicates a product operation, and $\forall \mathbf{e_{i}, e_{j}} \in \mathbf{E}$. $EP$ denotes element-wise product, while $IP$ represents vector inner product.

Thirdly, attention mechanisms have been widely employed in CTR  \cite{AFM, autoint, DIEN, DIN,fignn}, but these attention mechanisms are only feature-level and do not work well to weight the information in the semantic space. Therefore, we introduce a spatial-level attention mechanism to better aggregate information from various semantic spaces within the fusion layer, in which the spatial-level attention mechanism is:
\begin{equation}
\label{9}
\begin{aligned}
&\mathbf{K}_i = \textbf{MLP}_{k}^{i}\left(space_{i}\right), \\
\end{aligned}
\end{equation}
where $\textbf{MLP}_{k}$ and $\textbf{MLP}_{v}$ are nearly identical, with the difference being the use of LeakyReLU activation in the output layer instead of no activation function. $\mathbf{K}$ is a scalar that represents the weight of information implied by the current semantic $\mathbf{space}$ (e.g., $\mathbf{E}$ and $\mathbf{S}$). To facilitate the discussion, we collectively refer to the subcomponent consisting of $\textbf{MLP}_{k}$ and $\textbf{MLP}{v}$ as the \textit{Key-Value Block}.

In the end, we obtain the final prediction result by performing a straightforward weighted summation pooling as the fusion layer.
\begin{equation}
\begin{aligned}
&\hat{y} = \sum_{i=1}^H \mathbf{K}_i\mathbf{v}_i,
\end{aligned}
\end{equation}
where $H$ represents the number of semantic spaces. We call this model, which simply and efficiently captures feature interaction information from multi-semantic spaces, the \textbf{Simple Multi-Head Network (simMHN)}. This model is the predecessor of CETN, whose performance and architecture are shown in Table \ref{table 2} and Figure \ref{figure 4}.

It can be observed that this simple model achieves state-of-the-art performance, which obtains sub-optimal results and even outperforms the strongest baseline model FinalMLP \cite{finalmlp}.

Next, we attempt to enhance the model's performance by focusing on the diversity and homogeneity of information, without significantly altering the complexity of the simMHN model.

\subsection{How to Improve the Diversity of Information Captured}
\label{section 4.2}
Perturbation operations in contrastive learning can further differentiate information within semantic spaces  \cite{simGCL, sgl}. Therefore, we employ product \& perturbation as the better segmentation approach for semantic spaces and distinguish between primary and auxiliary semantic spaces (the semantic space processed by segmentation is called the auxiliary space, and vice versa). In this way we can get an augmented $\mathbf{E^\prime}$:
\begin{equation}
\label{11}
\begin{aligned}
&\mathbf{E^\prime}  =\mathbf{E}+\Delta^{\prime}, \\
&\Delta^{\prime}  =\bar{\Delta} \odot \operatorname{sign}\left(\mathbf{E}\right), \\ 
&\bar{\Delta} \in \mathbb{R}^d \sim U(0,1),
\end{aligned}
\end{equation}
where $\Delta^{\prime}$ represents a noise signal, 
$\bar{\Delta}$ is a sample drawn from a uniform distribution between 0 and 1. Afterwards, we further differentiate information within the semantic space using the product operation:
\begin{equation}
\label{12}
\begin{aligned}
\mathbf{S}_{EP}^\prime &= \Vert_{i=1}^n \Vert_{j=i}^n\ \mathcal{F}_{EP}(\mathbf{e_i^\prime} , \mathbf{e_j^\prime}), \\
\mathbf{S}_{IP}^\prime &= \Vert_{i=1}^n \Vert_{j=i}^n\ \mathcal{F}_{IP}(\mathbf{e_i^\prime} , \mathbf{e_j^\prime}),
\end{aligned}
\end{equation}
where $\forall \mathbf{e_i^\prime},  \mathbf{e_j^\prime} \in \mathbf{E^\prime}$. After segmentation, we improve the diversity of information within the semantic space. However, it becomes apparent that there is no inherent supervisory signal between multiple subcomponents to ensure their ability to capture distinct feature interactions. Relying solely on a self-adaptive joint learning strategy among subcomponents often leads to suboptimal performance. Therefore, we introduce auxiliary supervisory signals for the real-valued semantic space.

Given that we imitate the augmentation concept from contrastive learning during segmentation, we employ a contrastive loss to supervise the information-capturing behavior of subcomponents within the auxiliary semantic space. However, it's worth noting that the alignment strategy in contrastive learning is not always effective, as indicated in previous studies  \cite{simGCL,XsimGCL}. A similar observation holds for CTR tasks based on user historical behavior sequences  \cite{MISS}. As a result, we modify the contrastive loss, abandoning alignment while retaining uniformity. More specifically, we consider the information in both auxiliary spaces to be negative samples even if they are obtained from the same $\mathbf{S}$, and the modified loss is formulated to minimize the agreement with these samples, as follows:
\begin{equation}
\label{13}
   \mathcal{L}_{c l}=\sum_{i \in \mathcal{B}}-\log \frac{\exp \left(1 / \tau\right)}{\sum_{j \in \mathcal{B}} \exp \left(\text{sim}(\mathbf{V}_i^{\prime}, \mathbf{V}_j^{\prime \prime}) / \tau\right)},
\end{equation}
where $\mathbf{V^{\prime}}$, $\mathbf{V^{\prime\prime}} \in \mathbb{R}^{d_v}$ represent the real-values (obtained from different \textit{Key-Value Blocks}) within the auxiliary semantic space after segmentation, $d_v$ represents the real-values vector dimension. We refer to this modified loss as \textbf{Denominator-only InfoNCE} (Do-InfoNCE).

For subcomponents, we seek to further enhance their inherent ability to capture information from different semantic spaces. Some studies suggest that choosing appropriate activation functions can significantly impact the performance of neural networks in various application scenarios  \cite{activations1,activations2,activations3}. For instance, when there is excessive noise in the current semantic space, we can consider using ReLU as the activation function to filter out irrelevant information. Therefore, utilizing suitable activation functions in different semantic spaces can aid the model in effectively capturing diverse feature interactions. We employ different activation functions within $\textbf{MLP}_{v}$ belonging to different semantic spaces. 

\subsection{How to Ensure Homogeneity of Information}
After segmenting the semantic space and enhancing the ability of individual subcomponents to capture diverse information, while the model can better capture a variety of information, the increase in the number of feature spaces can lead to an issue of excessive noise. Therefore, we further introduce the concept of homogeneity, which complements diversity. Specifically, we aim for the information captured by various semantic spaces to be as distinct as possible (diversity). However, we also seek this information to be fundamentally similar (homogeneity), without being entirely dissimilar. To achieve this goal, we draw inspiration from residual networks  \cite{resNet} and introduce a \textbf{Through Network} to ensure the homogeneity of the captured information across various semantic spaces. Formally, we define this \textbf{Through Connection} as:
\begin{equation}
\label{14}
\begin{aligned}
   &\mathbf{V}\ =\textbf{MLP}_{v}(\mathbf{E}), \\   &\mathbf{V^\prime}=\textbf{MLP}_{v}^\prime(\mathbf{S}_{EP}^\prime) + \mathbf{V},\\
   &\mathbf{V^{\prime\prime}}=\textbf{MLP}_{v}^{\prime\prime}(\mathbf{S}_{IP}^\prime) + \mathbf{V},
\end{aligned}
\end{equation}
where $\mathbf{V} \in \mathbb{R}^{d_v}$ denotes the real-values information in the main semantic space, $\mathbf{S^\prime}$ represents the augmented embeddings in the auxiliary semantic space. By constructing the model in this manner, we establish a framework that is able to connect subcomponents in multiple semantic spaces, ensuring their homogeneity in capturing information while mitigating overfitting. Additionally, we also simply introduce the cosine similarity as an auxiliary loss, which takes the following form:
\begin{equation}
\label{15}
\begin{aligned}
   &\mathcal{L}^{\prime}_{cos}=\sum_{i \in \mathcal{B}} 1 - \text{sim}(\mathbf{V_{i}}, \mathbf{V^{\prime}_{i}}), \\
   &\text{sim}(\mathbf{V}, \mathbf{V^\prime}) = \frac{\mathbf{V}^{\top} \mathbf{V^\prime}}{\left\|\mathbf{V}\right\|\left\|\mathbf{V^\prime}\right\|}.
\end{aligned}
\end{equation}
The $\mathcal{L}^{\prime}_{cos}$ encourages homogeneity between $\mathbf{V}$ and $\mathbf{V^\prime}$, other similarity functions can also be utilized here.

\subsection{Fusion Layer and Multi-task Training}
After capturing information concurrently from $H$ semantic spaces, the real-values ($\mathbf{V}_i$) and weights ($\mathbf{K}_i$) from each semantic space are aggregated by the fusion layer:
\begin{equation}
\label{16}
\begin{aligned}
&\hat{y} = \sum_{i=1}^H \mathbf{K}_i(\mathbf{W}^{\top}\mathbf{V}_i + \mathbf{b}),
\end{aligned}
\end{equation}
where $\mathbf{W} \in \mathbb{R}^{d_v}$ and $\mathbf{b}$ are the weight and bias parameters. At this point, we have obtained a new CTR model, \textbf{Contrast-enhanced
Through Network (CETN)}, whose architecture is illustrated in Figure \ref{figure 3}.

As we are targeting the click-through rate, which is a binary classification task, we have employed the widely-used logloss  \cite{fignn,xdeepfm,openbenchmark,autoint} as the loss function for our model:
\begin{equation}
\label{18}
\mathcal{L}_{ctr}=-\frac{1}{N} \sum_{i=1}^{N}\left(y_{i} \log \left(\hat{y}_{i}\right)+\left(1-y_{i}\right) \log \left(1-\hat{y}_{i}\right)\right),
\end{equation}
where $N$ is the total number of training samples, $i$ represents the sample index, $y$ and $\hat{y}$ represent the true label and the predicted result of CETN. Next, we can obtain the total loss, denoted as $\mathcal{L}_{total}$:
\begin{equation}
\label{19}
\mathcal{L}_{total}=\mathcal{L}_{ctr} + \alpha \cdot \mathcal{L}_{c l} + (\beta^\prime \cdot \mathcal{L}^{\prime}_{cos} + \beta^{\prime\prime} \cdot \mathcal{L}^{\prime\prime}_{cos}),
\end{equation}
where $\alpha$, $\beta^\prime$, $\beta^{\prime\prime}$ represent hyperparameters that control the balance between the loss functions.

\subsection{Training Procedure}
In the above sections, we engaged in a detailed discussion on how to simply yet effectively capture feature interactions across various semantic spaces, as well as how to balance the diversity and homogeneity of the captured information. To facilitate a deeper understanding of our proposed CETN model, we provide a comprehensive depiction of its training process in Algorithm \ref{Algorithm 1}.

\begin{algorithm}[t]
    \caption{The overall training process of CETN}
    \label{Algorithm 1}
    \leftline{\quad \textbf{\ \  Require: } feature embeddings $\mathbf{E}$;} 
            \begin{algorithmic}[1]
            \STATE Initialize all parameters.\\
            \STATE \textbf{if} use perturbing \textbf{then}\\
            \STATE \quad Get $\mathbf{E}^{\prime}$ according to Equation (\ref{11});\\
            \STATE \textbf{else} \\
            \STATE \quad Get $\mathbf{E}^{\prime}$ = $\mathbf{E}$;\\
            \STATE \textbf{end} \\
            \STATE Construct $\mathbf{S}^{\prime}_{EP}$ and $\mathbf{S}^{\prime}_{IP}$ according to Equation (\ref{12});\\
            
            \STATE \textbf{while} CETN not converge \textbf{do} \\
            \STATE \quad Calculate spatial information weights $\mathbf{K}$, $\mathbf{K}^{\prime}$, and $\mathbf{K}^{\prime\prime}$ according to Equation (\ref{9});\\
            \STATE \quad Calculate real-values of each semantic space $\mathbf{V}$, $\mathbf{V}^{\prime}$, and $\mathbf{V}^{\prime\prime}$ according to Equation (\ref{14});\\
            \STATE \quad Calculate $\mathcal{L}_{cl}$ according to Equation (\ref{13});\\
           \STATE \quad Calculate $\mathcal{L}_{cos}$ according to Equation (\ref{15});\\
            \STATE \quad Fusion information from various semantic spaces according to Equation (\ref{16});
            \STATE \quad Update the model parameters using Equation (\ref{19});
            \STATE \textbf{end while} \\
            \end{algorithmic}
\end{algorithm}

Initially, we decide whether to perturb the embedding $\mathbf{E}$ for augmentation (lines 2-6), followed by conducting product-based augmentation (line 7) to distinguish the feature interaction information implied in different semantic spaces. At this point, we have made the necessary segmentation of the semantic spaces. Subsequently, we employ the \textit{Key-Value Block} to capture feature interaction information within each semantic space (lines 9-10) and calculate the self-supervised signal (lines 11-12). Finally, we aggregate the information from all semantic spaces in the fusion layer and update the parameters (lines 13-14).

\subsection{Model Analysis}
\subsubsection{Model Size} To effectively capture feature interaction information across various semantic spaces, CETN employs the \textit{Key-Value Block} and product \& perturbation. For ease of discussion, we regard $W_{\Psi}$ as the set of weights in the corresponding MLP and ignore the embedding parameters. In the \textit{Key-Value Block}, it can be simply viewed as six parallel MLPs, so its corresponding space complexity is \textit{O}(6$|W_{\Psi}|$). For the product \& perturbation operation, we further divide additional auxiliary semantic spaces, hence its corresponding space complexity is \textit{O}($df$ + $\frac{df(f+1)}{2}$ + $\frac{f(f+1)}{2}$). As the space complexity represented by the inner product $\frac{f(f+1)}{2}$ is of constant level, therefore, the space complexity of CETN is \textit{O}(6$|W_{\Psi}|$ + $df$ + $\frac{df(f+1)}{2}$). For more detailed information on the parameter size, refer to Tables \ref{table 4} and \ref{table 5}.

\subsubsection{Time Complexity} In a manner similar to the calculation of space complexity, the inference time of CETN is primarily due to the \textit{Key-Value Block} and product \& perturbation. Therefore, the time complexity during inference is \textit{O}(6$|W_{\Psi}|$ + 2$dfs$ + $\frac{df(f+1)}{2}$ + ${df(f+1)}$). It's worth mentioning that due to the relationship between the Hadamard product and the inner product (the latter being the sum of the former), in practical operations, we can optimize it as follows: \textit{O}(6$|W_{\Psi}|$ + 2$dfs$ + ${df(f+1)}$).

Furthermore, for a more detailed comparison, we present the time complexities in the training of DNN, DCN, FinalMLP, xDeepFM, AFN+, and our proposed CETN model in Table \ref{table 2.5}. We let $L$, $U$, and $M$ represent the number of explicit interaction layers, logarithmic neurons, and feature maps, respectively. $d_1$, $d_2$, $W_{gate}$, and $f_{s}$ represent the output dimension of the two MLPs in FinalMLP, the number of parameters in the gate unit, and the feature fields to be filtered, respectively. $(h_1, h_2)$ represent the outputs of the two feature interaction (FI) encoders in CL4CTR.

\begin{table*}[t]
\centering
\caption{Comparison of Analytical Time Complexity \\
$N \gg s > |W_{\Psi}| \approx |W_{gate}| > d_1 \approx d_2 \approx |h_1| > f \approx d \approx f_s \approx M \approx U$}
\label{table 2.5}
\resizebox{\linewidth}{!}{
\begin{tabular}{c|c|c|c|c}
\hline \textbf { Model } & \textbf { Embedding } & \textbf {Implicit interaction} & \textbf {Explicit interaction} & \textbf { Objective Function } \\
\hline \hline \text { DNN } & \textit{O}(2$dfs$) & \textit{O}($|W_{\Psi}|$) & -  & \textit{O}($N$)\\
\hline \text { DCN } & \textit{O}(2$dfs$) & \textit{O}($|W_{\Psi}|$) & \textit{O}($4dfL$) & \textit{O}($N$)\\
\hline \text { FinalMLP } & \textit{O}(2$dfs$ + 2$dsf_{s}$) & \textit{O}(2$|W_{\Psi}|$) & \textit{O}($2d_1d_2 + \text{2}|W_{gate}| + \text{2}df_s$) & \textit{O}($N$)\\
\hline \text { xDeepFM } & \textit{O}(2$dfs$) & \textit{O}($|W_{\Psi}|$) & \textit{O}($dfM(1 + ML))$) & \textit{O}($N$)\\
\hline \text { AFN+ } & \textit{O}(4$dfs$) & \textit{O}($|W_{\Psi}|$) & \textit{O}($2df(1 + U)$) & \textit{O}($N$)\\
\hline \text { CL4CTR } & \textit{O}(2$dfs$) & \textit{O}(3$|W_{\Psi}|$) & - & \textit{O}($N + N|h_1| + \frac{dfN(1 + N)}{2} + \frac{dfN^2(f - 1)}{2}$)\\
\hline \text { CETN } & \textit{O}(2$dfs$) & \textit{O}(6$|W_{\Psi}|$) & \textit{O}(${df(1+f)}$) & \textit{O}($N$ + 2$Nd_v$)\\
\hline
\end{tabular}}
\end{table*}

It can be concluded that CETN, compared to other models that can also capture feature interaction information in multiple semantic spaces, has a comparable time complexity. It is worth noting that, due to the use of contrastive loss in CETN, this leads to a longer training time. However, it does not affect the corresponding inference speed in actual applications. In the case of implicit interactions, although CETN requires six MLPs, the practical time complexity does not increase proportionally. This is because MLPs are parallel-friendly and simple yet effective, which ensures the time complexity remains manageable.

\subsubsection{Comparison with AFN+} The Adaptive Feature Network (AFN) \cite{AFN} is a solid mathematical model that cleverly applies logarithmic operation rules for adaptive order explicit feature interaction. Its enhanced version, AFN+, employs DNN and uses a Logarithmic Transformation (LT) Layer and MLP as parallel components to capture feature interaction information in two semantic spaces. Unlike previous works \cite{deepfm, dcn}, it neither shares an embedding layer nor uses gating units to segment the semantic space, instead simply maintaining separate embedding matrices for the two parallel components. However, maintaining two sets of embedding matrices is inefficient due to the majority of parameters and computational costs being taken up by the embedding operation (the performance of AFN+ will be demonstrated in Section \ref{section 5}). The LT layer in AFN+ uses logarithmic operations to simulate the Hadamard product, which can be formulated as follows:
\begin{equation}
\exp \left(\sum_{i=1}^m w_{i j} \ln \mathbf{e}_i\right)=\mathbf{e}_1^{w_{1 j}} \odot \mathbf{e}_2^{w_{2 j}} \odot \ldots \odot \mathbf{e}_m^{w_{m j}},
\end{equation}
where $w{i j}$ is the coefficient of the $j$-th neuron in the $i$-th feature field and $\odot$ denotes Hadamard product. It is not difficult to observe that when the model learns the correct value for $w_{i j}$, it can adaptively learn feature interactions of any order. However, since $w_{i j}$ is usually non-zero, the interaction learned by AFN is essentially an exponentially weighted, fixed full-order feature interaction. Interestingly, it is pointed out in some existing works that high-order feature interactions do not bring the expected performance improvement \cite{autofis, aim}, leading to subpar performance of AFN+. In CETN, we only use the Hadamard product to model second-order feature interactions, serving as an enhanced auxiliary semantic space, thus achieving better performance in both model complexity and performance metrics.

\subsubsection{Comparison with CL4CTR} CL4CTR \cite{CL4CTR} is the first to introduce the contrastive learning paradigm into feature interaction-based CTR prediction models. It uses Euclidean distance loss for feature alignment to make feature representations in the same feature field as similar as possible, and employs cosine loss for field uniformity to maximize the difference between feature representations in different feature fields. Additionally, it uses a perturbation operation to divide into two new feature interaction auxiliary spaces and compares the encoded feature interaction information through a feature interaction (FI) encoder. Moreover, the CL4CTR still uses Euclidean distance loss as a contrastive loss to minimize the encoding differences between the two FI encoders, as shown in Equation (\ref{21}), without introducing the widely acclaimed InfoNCE loss.
\begin{equation}
\label{21}
\mathcal{L}_{CL4CTR}=\frac{1}{B} \sum_{i=1}^B\left\|{h}_{i, 1}-{h}_{i, 2}\right\|_2^2,
\end{equation}
The fundamental difference between CETN and CL4CTR lies in their approach to handling information captured in the auxiliary semantic spaces. Where CL4CTR aims to make the captured information increasingly similar, CETN uses Do-InfoNCE to increase the diversity (i.e., dissimilarity) of captured information, while ensuring homogeneity (i.e., similarity) with the information in the original semantic space. Furthermore, as shown in Table \ref{table 2.5}, the time complexity of the contrastive loss in CL4CTR is very high, making it challenging to realistically apply in real-world production scenarios.

\section{EXPERIMENTS}
\label{section 5}
In this section, we provide a detailed account of our experimental setup and substantiate the superiority of CETN over other state-of-the-art models through a fair and extensive series of experiments. Subsequently, we conduct ablation experiments to investigate the impact of our configured hyperparameters on model performance and assess the rationale behind the presence of various modules.

\begin{table}[h]
\centering
\caption{Dataset statistics}
\label{dataset}
\begin{tabular}{ccccc} 
\toprule 
\textbf{Dataset} & \textbf{\#Instances} & \textbf{\#Fields} & \textbf{\#Features} & \textbf{\#Split} \\
\midrule 
\textbf{Avazu}   & 40M & 24  & 8.37M & 8:1:1\\
\textbf{Criteo}  & 46M  & 39 & 5.55M & 8:1:1\\
\textbf{MovieLens} & 2,006,859  & 3  & 90,445 & 7:2:1\\
\textbf{Frappe} & 288,609 & 10  & 5,382 & 7:2:1\\
\bottomrule
\end{tabular}
\label{table 3}
\end{table}

\subsection{Experimental Settings}
To ensure a fair comparison, we closely followed the settings of the   \cite{openbenchmark, finalmlp} work and selected the same CTR benchmark datasets originating from real production environments. Table \ref{table 3} below provides detailed information about these datasets.

\begin{itemize}
\item \textbf{Avazu}\footnote{\url{https://www.kaggle.com/c/avazu-ctr-prediction}}: This dataset contains 10 days of data on user clicks to ads while using mobile devices, as well as 15 explicit and 9 anonymous feature fields.
\item \textbf{Criteo}\footnote{\url{https://www.kaggle.com/c/criteo-display-ad-challenge}}: It is the well-known CTR benchmark dataset, which contains a 7-day stream of real data from Criteo, covering 39 anonymous feature fields.
\item \textbf{MovieLens}\footnote{\url{https://grouplens.org/datasets/movielens/}}: It consists of users' tagging records on movies. The datasets have been widely used in various research on recommender systems.
\item \textbf{Frappe}\footnote{\url{http://baltrunas.info/research-menu/frappe}}: It contains app usage logs from users under different contexts (e.g., daytime, location). The target value indicates whether the user has used the app under the context.
\end{itemize}

\textbf{Data preprocessing}: We follow the approach outlined in  \cite{openbenchmark}. For the Avazu dataset, we transform the timestamp field it contains into three new feature fields: hour, weekday, and weekend. For the Criteo dataset, we discretize the numerical feature fields by rounding down each numeric value $x$ to $\lfloor \log^2(x) \rfloor$ if $x > 2$, and $x = 1$ otherwise. For all datasets' categorical features, infrequent features (min\_count=2) are uniformly replaced with a default "OOV" token.

\subsubsection{Evaluation Metrics}
In order to compare the performance, we utilize two commonly used metrics in CTR models: AUC and logloss  \cite{deepfm, autoint, fignn, pnn1}.
\begin{itemize}
\item \textbf{AUC}: AUC stands for Area Under the ROC Curve. It measures the probability that a positive instance will be ranked higher than a randomly chosen negative one. A higher AUC indicates a better performance.
\item \textbf{Logloss}: logloss is the calculation result of Equation (\ref{18}). A lower logloss suggests a better capacity for fitting the data.
\end{itemize}

It's worth noting that even a slight improvement in AUC is meaningful in the context of CTR prediction tasks. Typically, CTR researchers consider improvements at the \textbf{0.001-level (0.1\%)} to be statistically significant, as often highlighted in existing literature  \cite{widedeep,deepfm,dcn,EDCN,CL4CTR, openbenchmark}. Following previous work  \cite{DIN, relaimpr, CL4CTR, masknet}, we further use RelaImpr to measure the relative improvement of AUC and Logloss, as defined:
\begin{equation}
\begin{aligned}
    \text{RelaImpr}_{AUC}&=\left(\frac{\text { AUC }(\text{target model})-0.5}{\text{AUC}(\text{base model})-0.5}-1\right) \times 100 \%, \\
    \text{RelaImpr}_{Logloss}&=\left(\frac{\text { Logloss }(\text{base model}) - \text{Logloss}(\text{target model})}{\text{Logloss}(\text{base model})}\right) \times 100 \%, \\
\end{aligned}
\end{equation}

\subsubsection{Baselines}\ We compared CETN with some classical state-of-the-art (SOTA) models. Given that deep CTR models often perform better, for models that have both non-DNN and DNN versions, we tend to choose the latter. The list of models we have chosen in chronological order of publication is as follows:
\begin{itemize}
\item \textbf{LR}  \cite{LR}: Logistic regression (LR) is a simple baseline model for CTR prediction.
\item \textbf{FM}  \cite{FM}: This model employs factorization techniques to address the challenge of learning on sparse datasets.
\item \textbf{DNN}  \cite{DNN}: This approach utilizes a feedforward neural network that takes a straightforward concatenation of feature embeddings as input.
\item \textbf{IPNN}  \cite{pnn1}: The model is an inner product-based feedforward neural network.
\item \textbf{Wide\ \&\ Deep}  \cite{widedeep}: It encompasses logistic regression (wide network) and the integration of a feedforward neural network (deep network).
\item \textbf{DeepFM}  \cite{deepfm}: This model combines FM and feedforward neural networks in parallel by sharing embedding layers.
\item \textbf{NFM}  \cite{NFM}: This method vertically combines FM and feed-forward neural networks through the Bi-interaction layer.
\item \textbf{AFM}  \cite{AFM}: The model incorporates an attention mechanism to discern the importance of different feature interactions.
\item \textbf{DCN}  \cite{dcn}: The model introduces the CrossNet that can explicitly model feature interactions, and integrate feedforward neural networks in parallel.
\item \textbf{xDeepFM}  \cite{xdeepfm}: Similar to DCN, this model introduces the Compressed Interaction Network (CIN), enhancing feature interaction from bit-wise to vector-wise.
\item \textbf{FiGNN}  \cite{fignn}: This model pioneers the use of graph neural networks to model feature interactions.
\item \textbf{AutoInt+}  \cite{autoint}: This model is the first to learn higher-order feature interactions using a multi-headed attention mechanism.
\item \textbf{AFN+}  \cite{AFN}: The model utilizes a logarithmic transformation layer to learn adaptive-order feature interactions.
\item \textbf{DCNv2}  \cite{dcnv2}: Expanding upon DCN, this model enhances the projection matrix's dimensionality and introduces a mixture of low-rank expert systems to optimize model inference speed.
\item \textbf{EDCN}  \cite{EDCN}: This model introduces both a bridge module and a regulation module, thereby enhancing the performance of the DCN model.
\item \textbf{MaskNet}  \cite{masknet}: This model utilizes the MaskBlock as its foundational structure. The introduced Instance-Guided Mask in the MaskBlock assists the model in acquiring high-quality information more effectively.
\item \textbf{GraphFM}  \cite{graphfm}: This model employs graph structure learning to address FM's limitations in selecting and learning appropriate higher-order feature interactions.
\item \textbf{FinalMLP}  \cite{finalmlp}: The model demonstrates the efficacy of the two-stream MLP model for implicit feature interaction learning.
\item \textbf{CL4CTR}  \cite{CL4CTR}: This model pioneers the integration of contrastive learning into CTR prediction based on feature interaction by introducing the concepts of feature alignment and field uniformity.
\item \textbf{EulerNet}  \cite{EulerNet}: 
This model employs Euler's formula to explicitly model feature interactions, integrating it with linear layers to adaptively learn feature interactions of the arbitrary order.
\end{itemize}

\begin{table*}[t]
\centering
\caption{Performance comparison of different models for CTR prediction. We highlight the top-5 best results in each dataset. "+": Integrating the original model with DNN networks.} 
\label{table 4}
\resizebox{\linewidth}{!}{
\begin{tabular}{cccccccc}
\hline
\multicolumn{1}{c|}{} &
  \multicolumn{1}{c|}{} &
  \multicolumn{3}{c|}{\textbf{Avazu}} &
  \multicolumn{3}{c}{\textbf{Criteo}} \\ \cline{3-8} 
\multicolumn{1}{c|}{\multirow{-2}{*}{\textbf{Year}}} &
  \multicolumn{1}{c|}{\multirow{-2}{*}{\textbf{Models}}} &
  \multicolumn{1}{c}{Logloss$\downarrow$} &
  \multicolumn{1}{c|}{AUC$\uparrow$} &
  \multicolumn{1}{c|}{\#Params} &
  \multicolumn{1}{|c}{Logloss$\downarrow$} &
  \multicolumn{1}{c|}{AUC$\uparrow$} &
  \multicolumn{1}{c}{\#Params} \\
  \hline
\multicolumn{1}{c|}{2007} &
  \multicolumn{1}{c|}{LR \cite{LR}} &
  0.381727 &
  \multicolumn{1}{c|}{0.777286} &
  \multicolumn{1}{c|}{3.7M} &
  0.456573 &
  \multicolumn{1}{c|}{0.793558} &
  \multicolumn{1}{c}{5.5M} \\
\multicolumn{1}{c|}{2010} &
  \multicolumn{1}{c|}{FM \cite{FM}} &
  { 0.376240} &
  \multicolumn{1}{c|}{{ 0.786085}} &
  \multicolumn{1}{c|}{78.7M} &
  0.444295 &
  \multicolumn{1}{c|}{0.807856} &
  \multicolumn{1}{c}{116.5M} \\
\multicolumn{1}{c|}{2016} &
  \multicolumn{1}{c|}{DNN \cite{DNN}} &
  0.372142 &
  \multicolumn{1}{c|}{0.792717} &
  \multicolumn{1}{c|}{78.5M} &
  0.438233 &
  \multicolumn{1}{c|}{0.813728} &
  \multicolumn{1}{c}{115.7M} \\
\multicolumn{1}{c|}{2016} &
  \multicolumn{1}{c|}{IPNN \cite{pnn1}} &
  \textbf{0.371156(2)} &
  \multicolumn{1}{c|}{\textbf{0.794330(2)}} &
  \multicolumn{1}{c|}{75.6M} &
  0.438217 &
  \multicolumn{1}{c|}{0.813945} &
  \multicolumn{1}{c}{114.6M} \\
\multicolumn{1}{c|}{2016} &
  \multicolumn{1}{c|}{Wide \& Deep \cite{widedeep}} &
  0.372015 &
  \multicolumn{1}{c|}{0.792982} &
  \multicolumn{1}{c|}{82.2M} &
  0.438110 &
  \multicolumn{1}{c|}{0.813814} &
  \multicolumn{1}{c}{120.4M} \\
\multicolumn{1}{c|}{2017} &
  \multicolumn{1}{c|}{DeepFM \cite{deepfm}} &
  \textbf{0.371890(5)} &
  \multicolumn{1}{c|}{0.793116} &
  \multicolumn{1}{c|}{82.4M} &
  \textbf{0.437676(3)} &
  \multicolumn{1}{c|}{\textbf{0.814200(5)}} &
  \multicolumn{1}{c}{117.6M} \\
\multicolumn{1}{c|}{2017} &
  \multicolumn{1}{c|}{NFM \cite{NFM}} &
  0.373843 &
  \multicolumn{1}{c|}{0.789997} &
  \multicolumn{1}{c|}{79.2M} &
  0.446362 &
  \multicolumn{1}{c|}{0.805424} &
  \multicolumn{1}{c}{118.6M} \\
\multicolumn{1}{c|}{2017} &
  \multicolumn{1}{c|}{AFM \cite{AFM}} &
  0.378901 &
  \multicolumn{1}{c|}{0.782119} &
  \multicolumn{1}{c|}{78.7M} &
  0.444468 &
  \multicolumn{1}{c|}{0.807100} &
  \multicolumn{1}{c}{116.5M} \\
\multicolumn{1}{c|}{2017} &
  \multicolumn{1}{c|}{DCN \cite{dcn}} &
  0.372353 &
  \multicolumn{1}{c|}{0.793142} &
  \multicolumn{1}{c|}{76.5M} &
  0.438091 &
  \multicolumn{1}{c|}{0.814103} &
  \multicolumn{1}{c}{132.5M} \\
\multicolumn{1}{c|}{2018} &
  \multicolumn{1}{c|}{xDeepFM \cite{xdeepfm}} &
  0.371944 &
  \multicolumn{1}{c|}{0.793121} &
  \multicolumn{1}{c|}{79.4M} &
  0.437820 &
  \multicolumn{1}{c|}{\textbf{0.814291(4)}} &
  \multicolumn{1}{c}{120.5M} \\
\multicolumn{1}{c|}{2019} &
  \multicolumn{1}{c|}{FiGNN \cite{fignn}} &
  0.374346 &
  \multicolumn{1}{c|}{0.789236} &
  \multicolumn{1}{c|}{75.0M} &
  0.438860 &
  \multicolumn{1}{c|}{0.813340} &
  \multicolumn{1}{c}{111.1M} \\
\multicolumn{1}{c|}{2019} &
  \multicolumn{1}{c|}{AutoInt+ \cite{autoint}} &
  0.372085 &
  \multicolumn{1}{c|}{0.792639} &
  \multicolumn{1}{c|}{77.5M} &
  0.438919 &
  \multicolumn{1}{c|}{0.812950} &
  \multicolumn{1}{c}{170.5M} \\
\multicolumn{1}{c|}{2020} &
  \multicolumn{1}{c|}{AFN+ \cite{AFN}} &
  0.372007 &
  \multicolumn{1}{c|}{\textbf{0.793513(5)}} &
  \multicolumn{1}{c|}{177.9M} &
  0.439244 &
  \multicolumn{1}{c|}{0.813098} &
  \multicolumn{1}{c}{226.3M} \\
\multicolumn{1}{c|}{2021} &
  \multicolumn{1}{c|}{DCNv2 \cite{dcnv2}} &
  0.372111 &
  \multicolumn{1}{c|}{0.793050} &
  \multicolumn{1}{c|}{77.2M} &
  0.438392 &
  \multicolumn{1}{c|}{0.813951} &
  \multicolumn{1}{c}{114.9M} \\
\multicolumn{1}{c|}{2021} &
  \multicolumn{1}{c|}{EDCN \cite{EDCN}} &
  \textbf{0.371627(4)} &
  \multicolumn{1}{c|}{\textbf{0.793874(3)}} &
  \multicolumn{1}{c|}{75.7M} &
  \textbf{0.437742(4)} &
  \multicolumn{1}{c|}{\textbf{0.814292(3)}} &
  \multicolumn{1}{c}{112.8M} \\
  \multicolumn{1}{c|}{2021} &
  \multicolumn{1}{c|}{MaskNet \cite{masknet}} &
  \textbf{0.371623(3)} &
  \multicolumn{1}{c|}{\textbf{0.793677(4)}} &
  \multicolumn{1}{c|}{77.8M} &
  0.439455 &
  \multicolumn{1}{c|}{0.812424} &
  \multicolumn{1}{c}{116.8M} \\
\multicolumn{1}{c|}{2022} &
  \multicolumn{1}{c|}{GraphFM \cite{graphfm}} &
  0.372646 &
  \multicolumn{1}{c|}{0.791912} &
  \multicolumn{1}{c|}{75.0M} &
  0.441261 &
  \multicolumn{1}{c|}{0.810482} &
  \multicolumn{1}{c}{111.0M} \\
\multicolumn{1}{c|}{2023} &
  \multicolumn{1}{c|}{FinalMLP \cite{finalmlp}} &
  0.372084 &
  \multicolumn{1}{c|}{0.793177} &
  \multicolumn{1}{c|}{80.0M} &
  \textbf{0.437631(2)} &
  \multicolumn{1}{c|}{\textbf{0.814472(2)}} & 
  \multicolumn{1}{c}{116.2M} \\ 
  \multicolumn{1}{c|}{2023} &
  \multicolumn{1}{c|}{CL4CTR \cite{CL4CTR}} &
  0.373158 &
  \multicolumn{1}{c|}{0.791745} &
  \multicolumn{1}{c|}{76.2M} &
  \textbf{0.437794(5)} &
  \multicolumn{1}{c|}{0.814159} &
  \multicolumn{1}{c}{112.5M} \\
  \multicolumn{1}{c|}{2023} &
  \multicolumn{1}{c|}{EulerNet \cite{EulerNet}} &
   0.376032 &
  \multicolumn{1}{c|}{0.792391} &
  \multicolumn{1}{c|}{75.6M} &
  0.442632 &
  \multicolumn{1}{c|}{0.811003} & 
  \multicolumn{1}{c}{113.4M} \\ \hline
\multicolumn{1}{c|}{ours} &
  \multicolumn{1}{c|}{CETN} &
  \textbf{0.370402(1)} &
  \multicolumn{1}{c|}{\textbf{0.796238(1)}} &
  \multicolumn{1}{c|}{85.1M} &
  \textbf{0.437319(1)} &
  \multicolumn{1}{c|}{\textbf{0.814804(1)}} &
  \multicolumn{1}{c}{140.1M} \\
  \hline
\end{tabular}}
\end{table*}

\subsubsection{\textbf{Implementation Details}}\ We implemented all models using Pytorch \cite{PYTORCH} and refer to existing works \cite{openbenchmark, FuxiCTR}. We employ the Adam optimizer \cite{adam} to optimize all models, with a default learning rate set to 0.001. For the sake of fair comparison, we set the embedding dimension to 20, and the batch size to 10,000 for all models. The hyperparameters of the baseline model were configured based on the \textit{optimal values} provided in  \cite{FuxiCTR,openbenchmark} and their original paper. To prevent overfitting, we employed early stopping with a patience value of 2. To facilitate reproducible research, we have open-sourced the code and running logs of CETN\footnote{\url{https://github.com/salmon1802/CETN}}.

\begin{table*}[t]
\centering
\caption{Performance comparison of different models for CTR prediction. We highlight the top-5 best results in each dataset. "+": Integrating the original model with DNN networks.} 
\label{table 5}
\resizebox{\linewidth}{!}{
\begin{tabular}{cccccccc}
\hline
\multicolumn{1}{c|}{} &
  \multicolumn{1}{c|}{} &
  \multicolumn{3}{c|}{\textbf{MovieLens}} &
  \multicolumn{3}{c}{\textbf{Frappe}} \\ \cline{3-8} 
\multicolumn{1}{c|}{\multirow{-2}{*}{\textbf{Year}}} &
  \multicolumn{1}{c|}{\multirow{-2}{*}{\textbf{Models}}} &
  \multicolumn{1}{c}{Logloss$\downarrow$} &
  \multicolumn{1}{c|}{AUC$\uparrow$} &
  \multicolumn{1}{c|}{\#Params} &
  \multicolumn{1}{|c}{Logloss$\downarrow$} &
  \multicolumn{1}{c|}{AUC$\uparrow$} &
  \multicolumn{1}{c}{\#Params} \\
  \hline
\multicolumn{1}{c|}{2007} &
  \multicolumn{1}{c|}{LR  \cite{LR}} &
   0.345537 &
  \multicolumn{1}{c|}{0.931499} &
  \multicolumn{1}{c|}{88597} &
  0.364208 &
  \multicolumn{1}{c|}{0.931528} &
  \multicolumn{1}{c}{5384} \\
\multicolumn{1}{c|}{2010} &
  \multicolumn{1}{c|}{FM  \cite{FM}} &
  0.276611 &
  \multicolumn{1}{c|}{0.942978} &
  \multicolumn{1}{c|}{1.8M} &
  0.193952 &
  \multicolumn{1}{c|}{0.969701} &
  \multicolumn{1}{c}{0.1M} \\
\multicolumn{1}{c|}{2016} &
  \multicolumn{1}{c|}{DNN  \cite{DNN}} &
  0.208941 &
  \multicolumn{1}{c|}{0.969276} &
  \multicolumn{1}{c|}{2.1M} &
  0.169329 &
  \multicolumn{1}{c|}{0.980640} &
  \multicolumn{1}{c}{0.5M} \\
\multicolumn{1}{c|}{2016} &
  \multicolumn{1}{c|}{IPNN \cite{pnn1}} &
  0.206157 &
  \multicolumn{1}{c|}{\textbf{0.970354(3)}} &
  \multicolumn{1}{c|}{2.1M} &
  \textbf{0.151626(5)} &
  \multicolumn{1}{c|}{\textbf{0.984333(4)}} &
  \multicolumn{1}{c}{0.5M} \\
\multicolumn{1}{c|}{2016} &
  \multicolumn{1}{c|}{Wide \& Deep \cite{widedeep}} &
  0.207965 &
  \multicolumn{1}{c|}{0.969820} &
  \multicolumn{1}{c|}{2.2M} &
  0.152807 &
  \multicolumn{1}{c|}{0.984018} &
  \multicolumn{1}{c}{0.5M} \\
\multicolumn{1}{c|}{2017} &
  \multicolumn{1}{c|}{DeepFM \cite{deepfm}} &
  \textbf{0.205239(4)} &
  \multicolumn{1}{c|}{0.969749} &
  \multicolumn{1}{c|}{2.2M} &
  0.153563 &
  \multicolumn{1}{c|}{0.983501} &
  \multicolumn{1}{c}{0.5M} \\
\multicolumn{1}{c|}{2017} &
  \multicolumn{1}{c|}{NFM \cite{NFM}} &
  0.302908 &
  \multicolumn{1}{c|}{0.939400} &
  \multicolumn{1}{c|}{2.1M} &
   0.160586 &
  \multicolumn{1}{c|}{0.980831} &
  \multicolumn{1}{c}{0.4M} \\
\multicolumn{1}{c|}{2017} &
  \multicolumn{1}{c|}{AFM \cite{AFM}} &
  0.277981 &
  \multicolumn{1}{c|}{0.942794} &
  \multicolumn{1}{c|}{1.8M} &
  0.242329 &
  \multicolumn{1}{c|}{0.955657} &
  \multicolumn{1}{c}{0.1M} \\
\multicolumn{1}{c|}{2017} &
  \multicolumn{1}{c|}{DCN \cite{dcn}} &
  0.204643 &
  \multicolumn{1}{c|}{\textbf{0.970140(5)}} &
  \multicolumn{1}{c|}{2.1M} &
  0.155340 &
  \multicolumn{1}{c|}{0.983995} &
  \multicolumn{1}{c}{0.5M} \\
\multicolumn{1}{c|}{2018} &
  \multicolumn{1}{c|}{xDeepFM \cite{xdeepfm}} &
   0.206501 &
  \multicolumn{1}{c|}{0.969769} &
  \multicolumn{1}{c|}{2.3M} &
  0.155676 &
  \multicolumn{1}{c|}{0.983409} &
  \multicolumn{1}{c}{0.5M} \\
\multicolumn{1}{c|}{2019} &
  \multicolumn{1}{c|}{FiGNN \cite{fignn}} &
 0.256296 &
  \multicolumn{1}{c|}{0.952781} &
  \multicolumn{1}{c|}{1.8M} &
  0.226627 &
  \multicolumn{1}{c|}{0.964828} &
  \multicolumn{1}{c}{0.2M} \\
\multicolumn{1}{c|}{2019} &
  \multicolumn{1}{c|}{AutoInt+ \cite{autoint}} &
  \textbf{0.203297(3)} &
  \multicolumn{1}{c|}{\textbf{0.970191(4)}} &
  \multicolumn{1}{c|}{2.2M} &
  \textbf{0.150433(2)} &
  \multicolumn{1}{c|}{\textbf{0.984076(5)}} &
  \multicolumn{1}{c}{0.7M} \\
\multicolumn{1}{c|}{2020} &
  \multicolumn{1}{c|}{AFN+ \cite{AFN}} &
  \textbf{0.205358(5)} &
  \multicolumn{1}{c|}{0.969492} &
  \multicolumn{1}{c|}{8.0M} &
  0.156007 &
  \multicolumn{1}{c|}{0.980949} &
  \multicolumn{1}{c}{3.8M} \\
\multicolumn{1}{c|}{2021} &
  \multicolumn{1}{c|}{DCNv2 \cite{dcnv2}} &
  0.206726 &
  \multicolumn{1}{c|}{0.969712} &
  \multicolumn{1}{c|}{2.1M} &
  \textbf{0.150948(4)} &
  \multicolumn{1}{c|}{\textbf{0.984368(3)}} &
  \multicolumn{1}{c}{0.6M} \\
\multicolumn{1}{c|}{2021} &
  \multicolumn{1}{c|}{EDCN \cite{EDCN}} &
  0.228215 &
  \multicolumn{1}{c|}{0.968716} &
  \multicolumn{1}{c|}{1.7M} &
  0.161859 &
  \multicolumn{1}{c|}{0.983283} &
  \multicolumn{1}{c}{0.2M} \\
\multicolumn{1}{c|}{2021} &
  \multicolumn{1}{c|}{MaskNet \cite{masknet}} &
  0.242475 &
  \multicolumn{1}{c|}{0.968499} &
  \multicolumn{1}{c|}{2.8M} &
  0.189036 &
  \multicolumn{1}{c|}{0.982834} &
  \multicolumn{1}{c}{1.6M} \\
\multicolumn{1}{c|}{2022} &
  \multicolumn{1}{c|}{GraphFM \cite{graphfm}} &
  0.222897 &
  \multicolumn{1}{c|}{0.964445} &
  \multicolumn{1}{c|}{1.7M} &
  0.288119 &
  \multicolumn{1}{c|}{0.939295} &
  \multicolumn{1}{c}{0.1M} \\
\multicolumn{1}{c|}{2023} &
  \multicolumn{1}{c|}{FinalMLP \cite{finalmlp}} &
  \textbf{0.196641(2)} &
  \multicolumn{1}{c|}{\textbf{0.972373(2)}} &
  \multicolumn{1}{c|}{2.2M} &
  \textbf{0.150553(3)} &
  \multicolumn{1}{c|}{\textbf{0.984854(2)}} & 
  \multicolumn{1}{c}{0.6M} \\ 
  \multicolumn{1}{c|}{2023} &
  \multicolumn{1}{c|}{CL4CTR \cite{CL4CTR}} &
 0.206971 &
  \multicolumn{1}{c|}{0.969982} &
  \multicolumn{1}{c|}{2.4M} &
  0.152105 &
  \multicolumn{1}{c|}{0.983712} &
  \multicolumn{1}{c}{1.0M} \\
  \multicolumn{1}{c|}{2023} &
  \multicolumn{1}{c|}{EulerNet \cite{EulerNet}} &
  0.213534 &
  \multicolumn{1}{c|}{0.965486} &
  \multicolumn{1}{c|}{1.7M} &
  0.153704 &
  \multicolumn{1}{c|}{0.980542} &
  \multicolumn{1}{c}{0.2M} \\ \hline
\multicolumn{1}{c|}{ours} &
  \multicolumn{1}{c|}{CETN} &
  \textbf{0.185652(1)} &
  \multicolumn{1}{c|}{\textbf{0.973957(1)}} &
  \multicolumn{1}{c|}{1.9M} &
  \textbf{0.150283(1)} &
  \multicolumn{1}{c|}{\textbf{0.985710(1)}} &
  \multicolumn{1}{c}{1.6M} \\
  \hline
\end{tabular}}
\end{table*}

\subsection{Overall Comparison}
The performance of CETN and the baseline model is shown in Table \ref{table 4} and Table \ref{table 5}. 
it can be observed that deep Click-Through Rate (CTR) models, with DeepFM  \cite{deepfm} as their representative, consistently outperform shallow models, typified by FM  \cite{FM}. This underscores the effectiveness of combining implicit high-order feature interactions with explicit shallow feature interactions, resulting in improved click-through rate predictions. Concurrently, it emphasizes the necessity of leveraging explicit feature interactions to overcome the performance bottlenecks associated with MLP.

\begin{table*}[t]
\centering
\caption{Relative improvement of AUC and Logloss with CETN. $\triangle \text{Logloss}$ and $\triangle \text{AUC}$ denote the average performance improvement.}
\label{table 6}
\resizebox{\linewidth}{!}{
\begin{tabular}{c|cc|cc|cc|cc|cc}
\hline
\multirow{2}{*}{\textbf{Model}} & \multicolumn{2}{c|}{\textbf{Avazu$_{RelaImpr}$}}          & \multicolumn{2}{c|}{{\textbf{Criteo$_{RelaImpr}$}}} & \multicolumn{2}{c|}{{\textbf{MovieLens$_{RelaImpr}$}}}      & \multicolumn{2}{c|}{{\textbf{Frappe$_{RelaImpr}$}}} & \multirow{2}{*}{$\triangle \text{Logloss}\downarrow$} & \multirow{2}{*}{$\triangle \text{AUC}\uparrow$} \\ \cline{2-9}
& Logloss$\downarrow$ & AUC(\%)$\uparrow$   & Logloss$\downarrow$ &AUC(\%)$\uparrow$      & Logloss$\downarrow$ & AUC(\%)$\uparrow$   & Logloss$\downarrow$       & AUC(\%)$\uparrow$  &    &     \\ \hline
LR  \cite{LR} & 2.97\% & 6.83\% & 4.22\% & 7.24\% &46.27\% & 9.84\% & 58.74\% & 12.56\% & 28.05\% & 8.44\%  \\
FM  \cite{FM} & 1.55\% & 3.55\% & 1.57\% & 2.26\% & 32.88\% & 6.99\% & 22.52\% & 3.41\%  & 14.63\%  & 4.95\%  \\ 
DNN  \cite{DNN} & 0.47\% & 1.20\% & 0.21\% & 0.34\% & 11.15\% & 1.00\% & 11.25\% & 1.05\%       & 5.77\% & 0.89\%  \\
IPNN  \cite{pnn1} & 0.20\% & 0.65\% & 0.20\% & 0.27\% & 9.95\% & 0.77\% & 0.89\% & 0.28\%     & 2.81\%  & 0.61\%  \\ 
Wide \& Deep \cite{widedeep} & 0.43\% & 1.11\% & 0.18\% & 0.32\% & 10.73\% & 0.88\% & 1.65\% & 0.35\%     & 3.25\%  &  0.80\% \\ 
DeepFM  \cite{deepfm} & 0.40\% & 1.07\% & 0.08\% & 0.19\% & 9.54\% & 0.90\% & 2.14\% & 0.46\%       & 3.04\% & 0.76\%  \\
NFM  \cite{NFM} & 0.92\% & 2.15\% & 2.03\% & 3.07\% & 38.71\% & 7.86\% & 6.42\% & 1.01\% & 12.02\%  & 5.24\%  \\ 
AFM   \cite{AFM} & 2.24\% & 5.00\% & 1.61\% & 2.51\% & 33.21\% & 37.04\% & 37.98\% & 6.60\% & 18.76\%  & 5.40\%  \\
DCN   \cite{dcn} & 0.52\% & 1.06\% & 0.18\% & 0.22\% & 9.28\% & 0.81\% & 3.26\% & 0.35\% & 3.31\%  & 0.73\%  \\
xDeepFM  \cite{xdeepfm} & 0.41\% & 1.06\% & 0.11\% & 0.16\% & 10.10\% & 0.89\% & 3.46\% & 0.48\%       & 3.52\% & 0.75\%  \\
FiGNN  \cite{fignn} & 1.05\% & 2.42\% & 0.35\% & 0.47\% & 27.56\% & 4.68\% & 33.69\% & 4.49\%     & 15.66\%  &  3.06\% \\ 
AutoInt+  \cite{autoint} & 0.45\% & 1.23\% & 0.36\% & 0.59\% & 8.68\% & 0.80\% & 0.10\% & 0.34\%       & 2.40\% & 0.86\%  \\
AFN+  \cite{AFN} & 0.43\% & 0.93\% & 0.44\% & 0.54\% & 9.60\% & 0.95\% & 3.67\% & 0.99\%  & 3.53\%  & 0.84\%  \\ 
DCNv2  \cite{dcnv2} & 0.46\% & 1.09\% & 0.24\% & 0.27\% & 10.19\% & 0.90\% & 0.44\% & 0.28\% & 2.83\% & 0.79\%  \\
EDCN  \cite{EDCN} & 0.33\% & 0.80\% & 0.10\% & 0.16\% & 18.65\% & 1.12\% & 7.15\% & 0.50\% & 6.56\% & 0.80\%  \\
MaskNet  \cite{masknet} & 0.33\% & 0.87\% & 0.49\% & 0.76\% & 23.43\% & 1.16\% & 20.50\% & 0.60\% & 11.19\% & 0.99\%  \\
GraphFM  \cite{graphfm} & 0.60\% & 1.48\% & 0.89\% & 1.39\% & 16.71\% & 2.05\% & 47.84\% & 10.57\% & 16.51\% & 1.74\%  \\
FinalMLP  \cite{finalmlp} & 0.45\% & 1.04\% & 0.07\% & 0.11\% & 5.59\% & 0.34\% & 0.18\% & 0.18\% & 1.57\% & 0.46\%  \\
CL4CTR  \cite{CL4CTR} & 0.74\% & 1.54\% & 0.11\% & 0.21\% & 10.30\% & 0.85\% & 1.20\% & 0.41\% & 3.09\% & 0.86\%  \\
EulerNet  \cite{EulerNet} & 1.50\% & 1.32\% & 1.20\% & 1.22\% & 13.06\% & 1.82\% & 2.23\% & 1.08\% & 4.50\% & 1.54\%  \\ \hline
\end{tabular}}
\vspace{-0.5em}
\end{table*}

Focusing on the models that achieved top-5 performance in the experiments, it can be observed that all of these models are based on parallel or stacked structures. Interestingly, over time, when all models are configured with their respective optimal parameters, the performance improvements of these SOTA models are relatively modest. Notably, the FinalMLP \cite{finalmlp}, consistently delivers strong performance across multiple datasets, further highlighting the importance of subcomponents and segments.

Taking an overall view, the CETN consistently maintains the highest AUC and Logloss performance, even when all models are configured with their respective optimal parameters. As is shown in Table \ref{table 6}, in comparison to the best baseline model, CETN exhibits improvements of 0.65\% (compared to IPNN), 0.11\% (compared to FinalMLP), 0.34\% (compared to FinalMLP), and 0.18\% (compared to FinalMLP) in AUC on the four datasets, while achieving corresponding reductions in Logloss by 0.2\% (compared to IPNN), 0.07\% (compared to FinalMLP), 5.59\% (compared to FinalMLP), and 0.1\% (compared to AutoInt+). Compared to the xDeepFM model, which also utilizes three semantic spaces, CETN achieves AUC average improvements of 0.75\% in AUC and 3.52\% in Logloss. This performance enhancement can be attributed to its ability to capture diversity within the feature space while preserving information homogeneity. It is noteworthy that, in contrast to intricate explicit feature interaction networks, the proposed CETN model simply integrates contrastive learning into simMHN, effectively improving its ability to capture feature interaction information.

Compared to the CL4CTR model, which also employs the contrastive learning paradigm, CETN demonstrates an average relative improvement of 3.09\% in Logloss and 0.86\% in AUC across the four datasets. Interestingly, in CL4CTR, the contrastive loss does not adopt the popular InfoNCE paradigm used in graph neural network-based recommender. Instead, it utilizes Euclidean distance loss as the contrastive loss, resulting in the maximization of similarity in the enhanced auxiliary semantic space embeddings. This stands in contrast to our Do-InfoNCE approach and is similar to $\mathcal{L}_{cos}$. This suggests that CL4CTR overlooks the connection between information in the auxiliary semantic space and the primary semantic space, emphasizing the necessity of homogeneity and diversity as unsupervised guiding signals for model learning. 

\subsection{Ablation Study}
\label{subsection 5.3}
In this section, we conducted extensive ablation studies to assess the contributions of individual modules of CETN to its overall performance. Several variants were designed to validate the effectiveness of the various CETN modules:
\begin{itemize}
\item \textbf{CETN (-A)}: To further enhance the diversity of information captured by the model, we employ different activation functions in multiple semantic spaces. To assess the necessity of this approach, we only specify the use of ReLU activation functions in the Multi-Layer Perceptron (MLP) for modeling feature interactions within multiple semantic spaces. This will reduce the diversity of information captured by the model.
\item \textbf{CETN (-CL)}: Contrastive loss serves to self-supervise and augment the diversity of information captured by the model. To determine whether it aids in improving model performance, we experiment by removing it from the model.
\item \textbf{CETN (-COS)}: The cosine loss primarily reinforces the homogeneity of the model from a supervised signal perspective. To ascertain its necessity, we exclude it from the model.
\item \textbf{CETN (-K)}: It serves to fine-tune the contribution weights of various subcomponents to the final prediction, further refining the model's predictive results. To establish the necessity of the $\textbf{MLP}_{k}$ (spatial-level attention) within the \textit{Key-Value Block}, we eliminate it from the model. 
\item \textbf{CETN (-P)}: To verify the effectiveness of our proposed product \& perturbation method, we replace it with the original embeddings $\mathbf{E}$ without any additional processing.
\item \textbf{CETN (-T)}: The Through Connection ensures the model maintains homogeneity in the information captured across multiple semantic spaces. To evaluate its usefulness, we remove it from the model.
\end{itemize}

\begin{figure*}[t]
    \begin{minipage}[t]{0.5\linewidth}
        \centering
        \includegraphics[width=\textwidth]{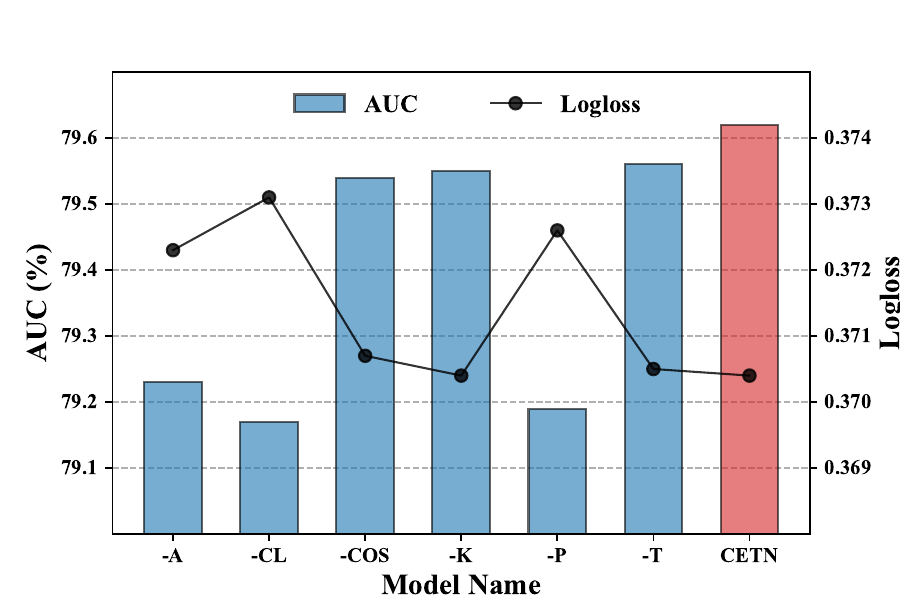}
        \centerline{(a) Avazu}
    \end{minipage}%
    \begin{minipage}[t]{0.5\linewidth}
        \centering
        \includegraphics[width=\textwidth]{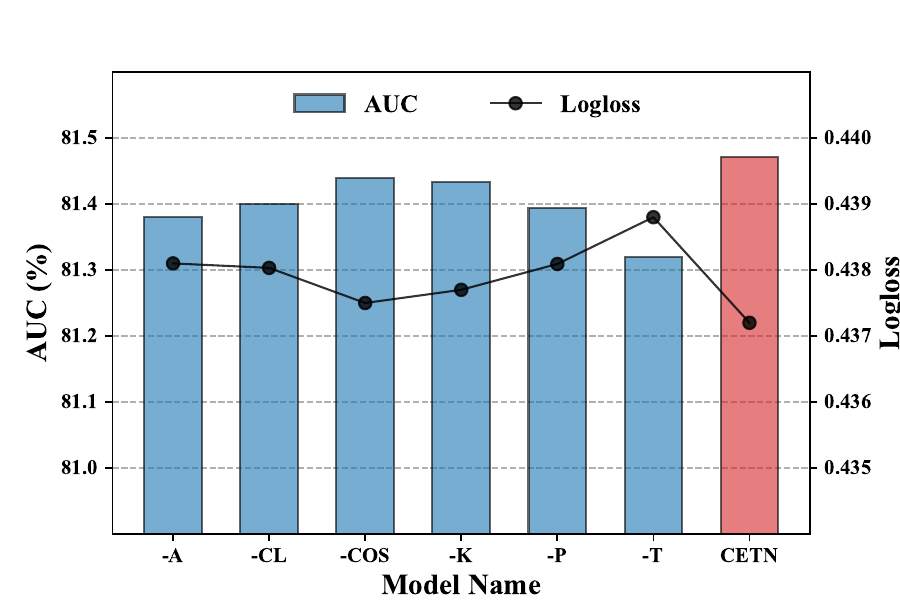}
        \centerline{(b) Criteo}
    \end{minipage}
    \begin{minipage}[t]{0.495\linewidth}
        \centering
        \includegraphics[width=\textwidth]{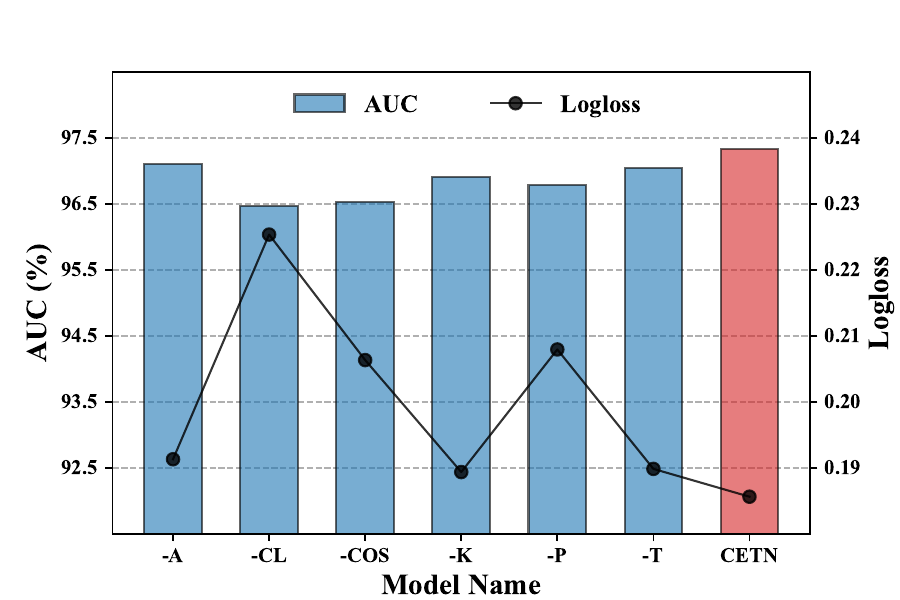}
        \centerline{(c) MovieLens}
    \end{minipage}
    \begin{minipage}[t]{0.495\linewidth}
        \centering
        \includegraphics[width=\textwidth]{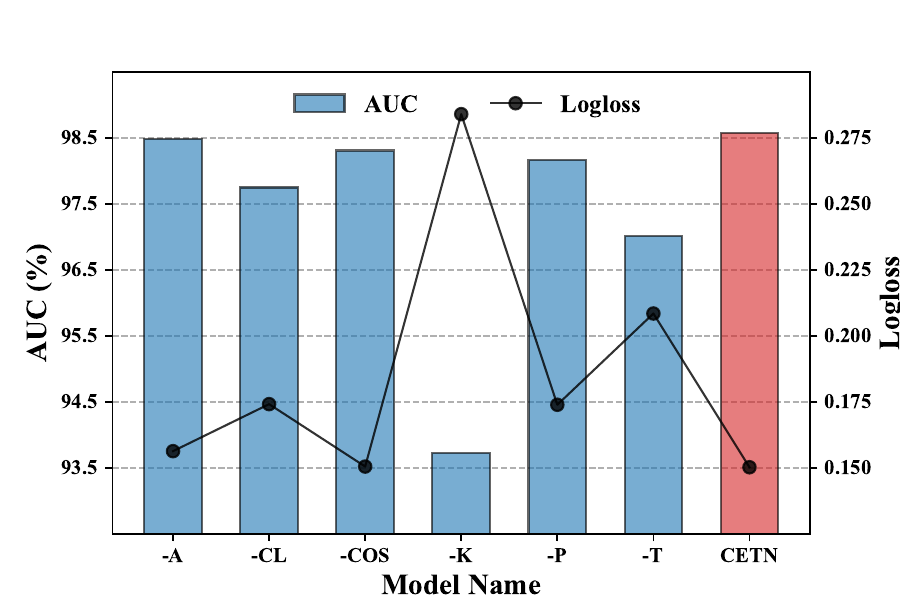}
        \centerline{(d) Frappe}
    \end{minipage}
    \captionsetup{justification=raggedright}
    \caption{Ablation study of CETN on Avazu (a), Criteo (b), MovieLens (c), and Frappe (d) datasets.}
    \label{figure 5}
\end{figure*}

Figure \ref{figure 5} illustrates the performance of CETN and its six variants. We can observe that CETN outperforms all the ablation models, providing compelling evidence for the necessity of each component in CETN. To delve deeper into this, let's break down the performance across different datasets.

Specifically, on the Avazu dataset, we notice that the performance drop is most pronounced for CETN (-A, -CL, -P). This indicates that the Avazu dataset benefits significantly from diverse feature interaction information to boost model performance. On the other hand, the lower performance loss of CETN (-COS, -T) suggests that the Avazu dataset is more in need of diverse feature interaction information to enhance model performance.

On the Criteo dataset, we observe the most significant performance drop in CETN (-T). This highlights that an increase in the number of feature fields often introduces more noise signals. Therefore, it becomes crucial to ensure information homogeneity while enhancing information diversity.

On the Movielens dataset, We empirically think that due to its limited number of feature fields and relatively fewer instances, it's susceptible to overfitting. Consequently, CETN (-CL, -COS), both complementing each other, play a significant role in constraining the scope of information captured by the model, allowing it to capture high-quality interaction information, thereby preventing overfitting to some extent. The correctness of this hypothesis can be intuitively observed in Figure \ref{figure 5} (c), where the model's performance experiences a noticeable decline when CETN removes $\mathcal{L}_{cl}$, $\mathcal{L}_{cos}$.

On the Frappe dataset, the performance of CETN (-K) exhibits a cliff-like decline, providing evidence for the effectiveness of the spatial-level attention mechanism. It is noteworthy that the Frappe dataset has only 0.7\% and 0.6\% of the data volume compared to the Avazu and Criteo datasets, respectively. This effectively underscores the challenge of models to adaptively assess the importance of information in various semantic spaces when dealing with relatively fewer data. Consequently, this leads to a poorer predictive performance of the model.

\begin{table}[ht]
\centering
\caption{Single performance of each semantic space. RelaImpr denotes the relative improvements compared with the simMHN. The underscore indicates the performance in the optimal semantic space.}
\resizebox{1.0\linewidth}{!}{
\begin{tabular}{c|cc|cc|cc|cc}
\hline
\multirow{2}{*}{Model} & \multicolumn{2}{c|}{Avazu} & \multicolumn{2}{c|}{Criteo} & \multicolumn{2}{c|}{MovieLens} & \multicolumn{2}{c}{Frappe} \\ \cline{2-9} 
                       & Logloss        & AUC       & Logloss        & AUC        & Logloss          & AUC         & Logloss        & AUC       \\ \hline
$\mathbf{S}_{EP}$   &    \underline{0.370488}            &    \underline{0.795332}       &     0.441558           &   0.811001         &   0.236711     &    0.964678         &     0.166236     &  \underline{0.982632}        \\
$\mathbf{S}_{IP}$  &   0.373471             &    0.790807       &     0.439198           &    0.813027        &     0.227524     &    0.960124         &      \underline{0.161219}         &    0.977373       \\
$\mathbf{E}$   &    0.372142            &    0.792717       &    \underline{0.438233}            &    \underline{0.813728}       &     \underline{0.208941}             &    \underline{0.969276}         &      0.169329          &   0.980640        \\ \hline
simMHN   &    0.371091    &    0.794810       &    0.437681            &    0.814355       &     0.199238             &    0.970165         &    0.180469            &    0.983528       \\
CETN   &   \textbf{0.370402}       &    \textbf{0.796238}       &    \textbf{0.437319}            &    \textbf{0.814804}       &     \textbf{0.185652}             &    \textbf{0.973957}         &      \textbf{0.150283}          &   \textbf{0.985710}        \\ \hline
RelaImpr   &   0.18\%       &    0.48\%       &    0.08\%            &    0.14\%       &    6.82\%              &   0.81\%         &      16.72\%          &   0.45\%        \\ \hline
\end{tabular}}
\label{table 7}
\end{table}

\subsection{Which Semantic Space is More Useful}
For the three semantic Spaces we define and their subcomponents, we conduct experiments to explore their individual contributions to the model's final predictions. The performance of each separate subcomponent on the four datasets is presented in Table \ref{table 7}. On the Avazu dataset, $\mathbf{S}_{EP}$ achieved outstanding performance, even surpassing IPNN (the best baseline model). However, its performance declined under the simple fusion of simMHN, demonstrating the drawbacks of this simple fusion approach. On the Criteo and MovieLens datasets, the original embeddings $\mathbf{E}$ showed better results, and simMHN further improved performance through simple fusion. On the Frappe dataset, both $\mathbf{S}_{EP}$ and $\mathbf{S}_{IP}$ respectively achieved the best AUC and Logloss, but in simMHN, although the AUC performance improved further, Logloss increased, indicating that simMHN might struggle to effectively predict true click probabilities. 

To further enhance the performance of the simMHN model without significantly increasing its complexity, we introduced diversity- and homogeneity-guided self-supervised signals. Additionally, we incorporated skip connections and various activation functions to balance both homogeneity and diversity. Consequently, CETN achieved improvements in Logloss by 0.18\%, 0.08\%, 6.82\%, and 16.72\% on the four datasets, and in AUC by 0.48\%, 0.14\%, 0.81\%, and 0.45\%, respectively. This demonstrates the effectiveness of ensuring both homogeneity and diversity in the information captured by the model.

\begin{figure*}[ht]
    \begin{minipage}[t]{0.245\linewidth}
        \centering
        \includegraphics[width=\textwidth]{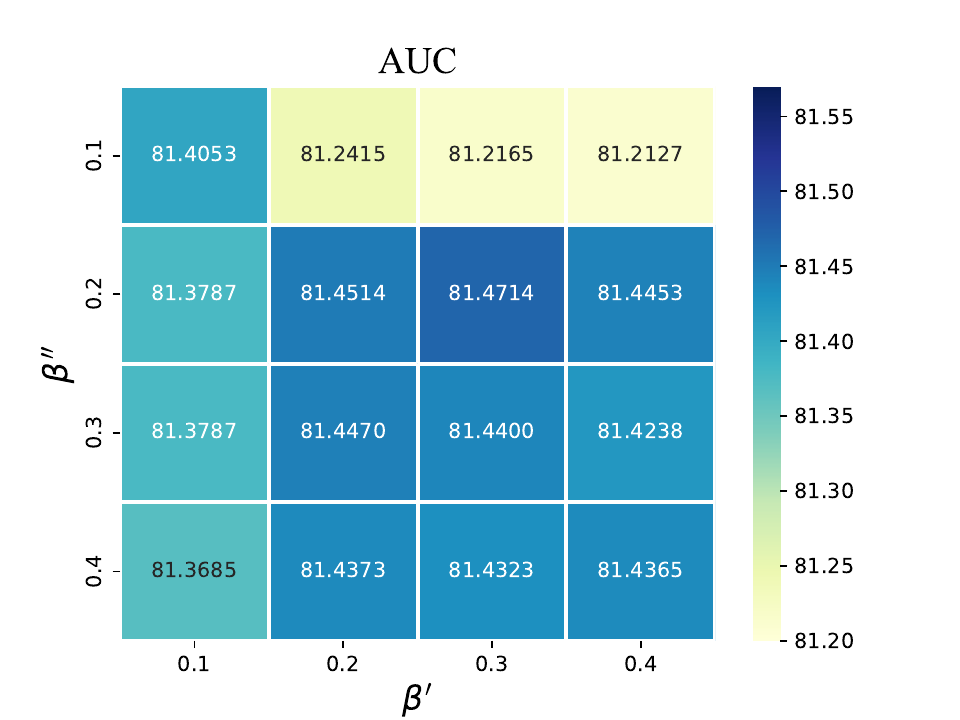}
        \centerline{(a) Criteo}
    \end{minipage}%
    \begin{minipage}[t]{0.245\linewidth}
        \centering
        \includegraphics[width=\textwidth]{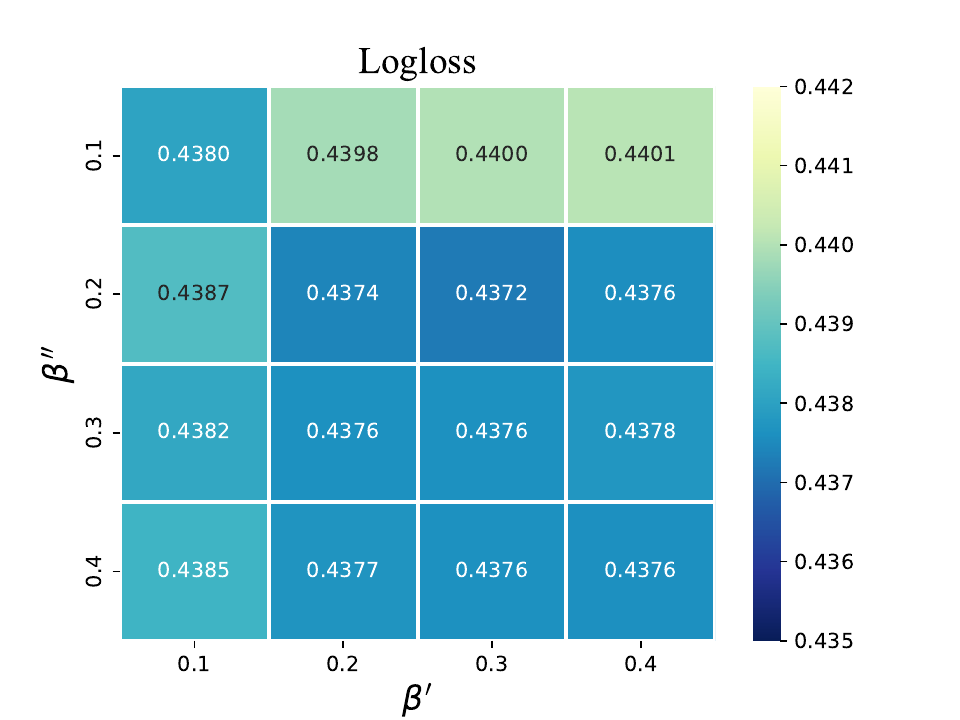}
        \centerline{(b) Criteo}
    \end{minipage}
    \begin{minipage}[t]{0.245\linewidth}
        \centering
        \includegraphics[width=\textwidth]{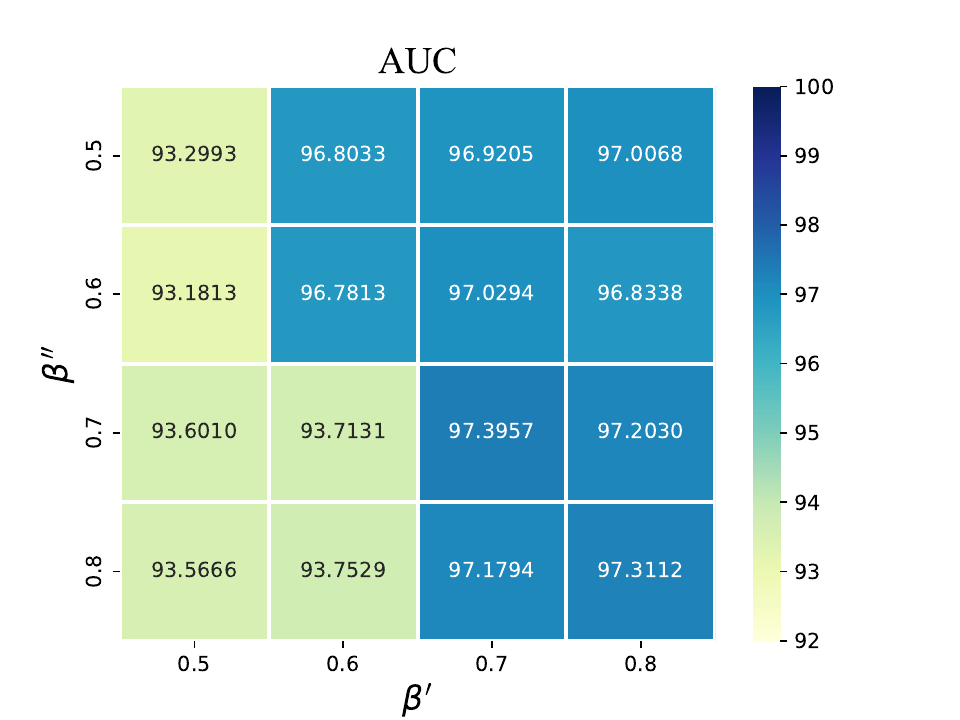}
        \centerline{(c) MovieLens}
    \end{minipage}%
    \begin{minipage}[t]{0.245\linewidth}
        \centering
        \includegraphics[width=\textwidth]{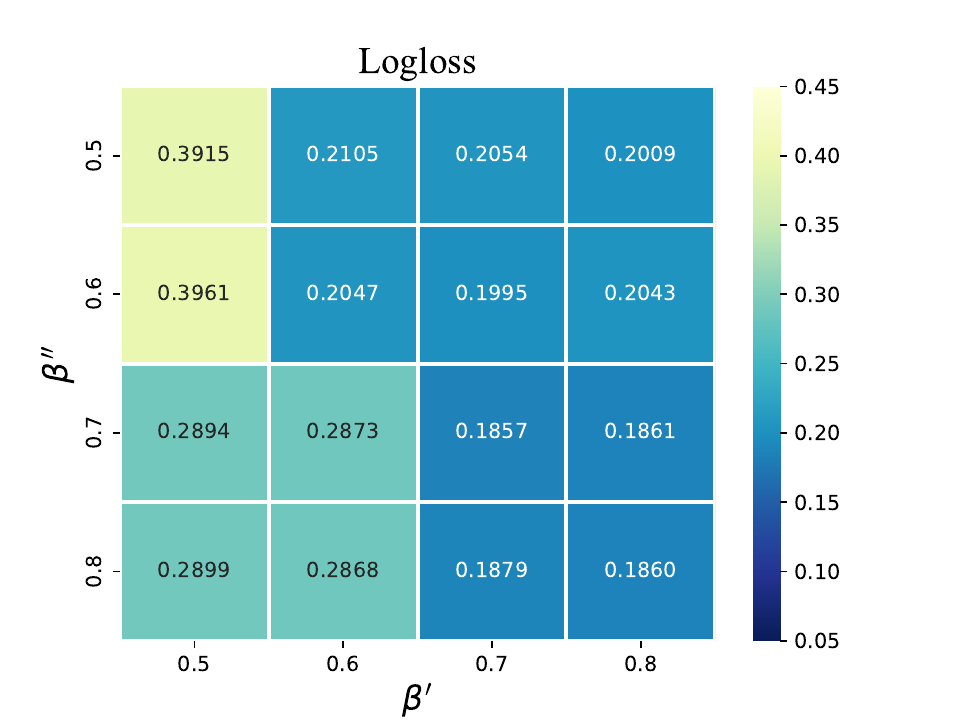}
        \centerline{(d) MovieLens}
    \end{minipage}
    \begin{minipage}[t]{0.5\linewidth}
        \centering
        \includegraphics[width=\textwidth]{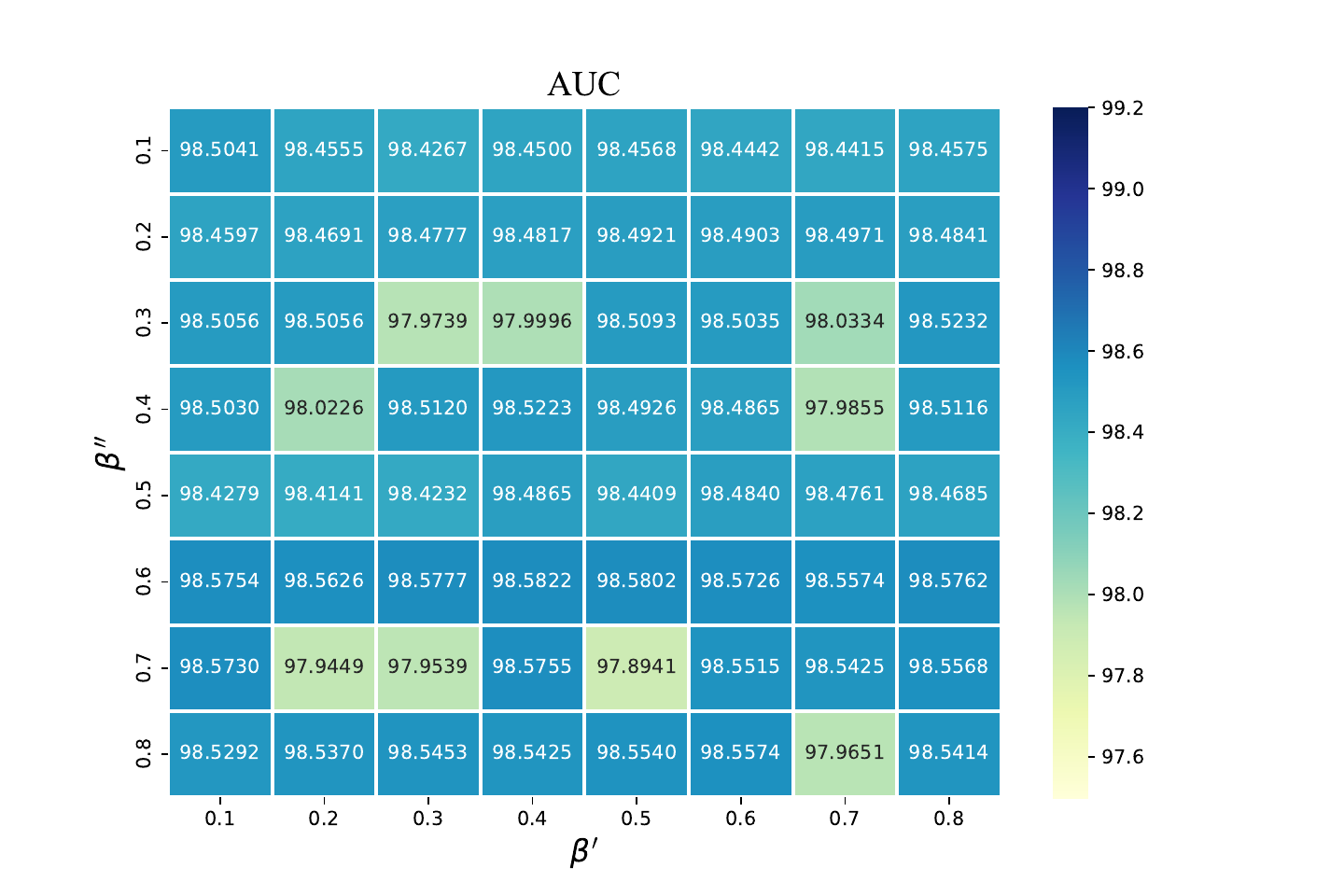}
        \centerline{(e) Frappe}
    \end{minipage}%
    \begin{minipage}[t]{0.5\linewidth}
        \centering
        \includegraphics[width=\textwidth]{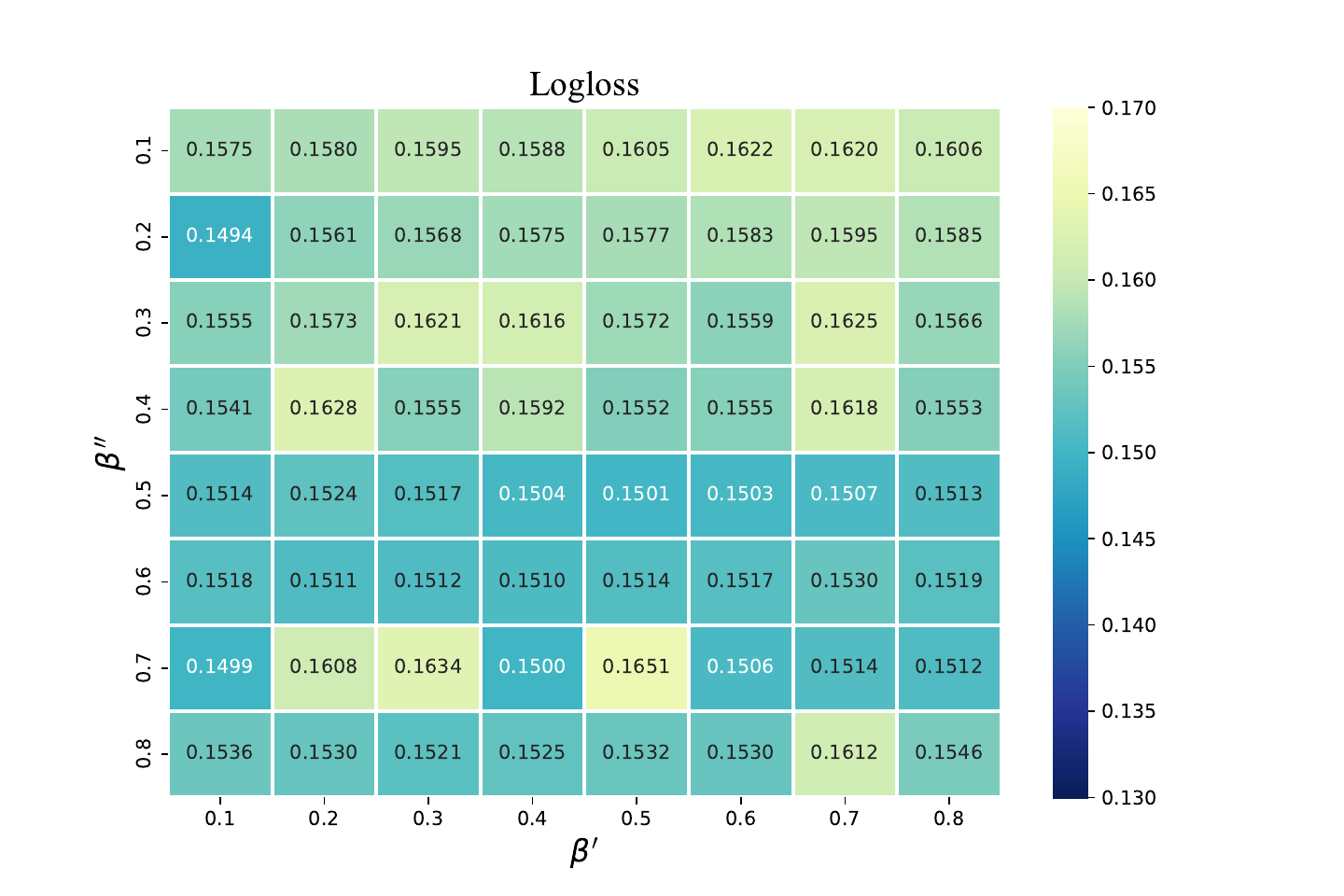}
        \centerline{(f) Frappe}
    \end{minipage}
    \captionsetup{justification=raggedright}
    \caption{\textbf{Performance comparison of different weights of cosine loss on Criteo (a,b), MovieLens (c,d), and Frappe (e,f) datasets. }}
    \label{figure 6}
\end{figure*}

\subsection{Hyper-parameter Analysis}
\subsubsection{Impact of the Weights in the $\mathcal{L}_{cos}$} We conduct a further investigation into the impact of different weight parameter combinations on the model's performance. For the sake of discussion, we confine the ranges of $\beta^\prime$ and $\beta^{\prime\prime}$ to [0.1 $\sim$ 0.4, 0.5 $\sim$ 0.8] while keeping other settings fixed. The results are presented as a heat map in Fig, \ref{figure 6}, and it is evident that CETN achieves its optimal performance when $\beta^\prime=0.3$ and $\beta^{\prime\prime}=0.2$ on the Criteo dataset. In the heatmap of MovieLens, multiple chunked regions in the model's performance are observed. The performance continues to decrease when $\beta^\prime$ and $\beta^{\prime\prime}$ are set to 0.5 or 0.6, but reaches optimal performance at 0.7 and 0.8. This phenomenon precisely validates the results presented in Table \ref{table 7}. Due to the relatively lower effective information content in the auxiliary semantic space, setting a larger $\mathcal{L}_{cos}$ becomes necessary to prevent the model from capturing noise and ensure homogeneity of information. 
On the Frappe dataset, it is evident that the model performs well when $\beta^{\prime\prime}$ is 0.6 $\sim$ 0.8 and achieves optimal results when $\beta^{\prime}$ equals 0.1, aligning with our previous reporting in Table \ref{table 7}. Notably, the model attains the lowest Logloss when $\beta^\prime$=0.1 and $\beta^{\prime\prime}$=0.2. This is reasonable, as on the Frappe dataset, even though $\mathbf{S}_{IP}$ exhibits poor AUC performance, which achieves the lowest Logloss in multiple semantic spaces.

\begin{figure*}[t]
    \begin{minipage}[t]{0.475\linewidth}
        \centering
    \includegraphics[width=\textwidth]{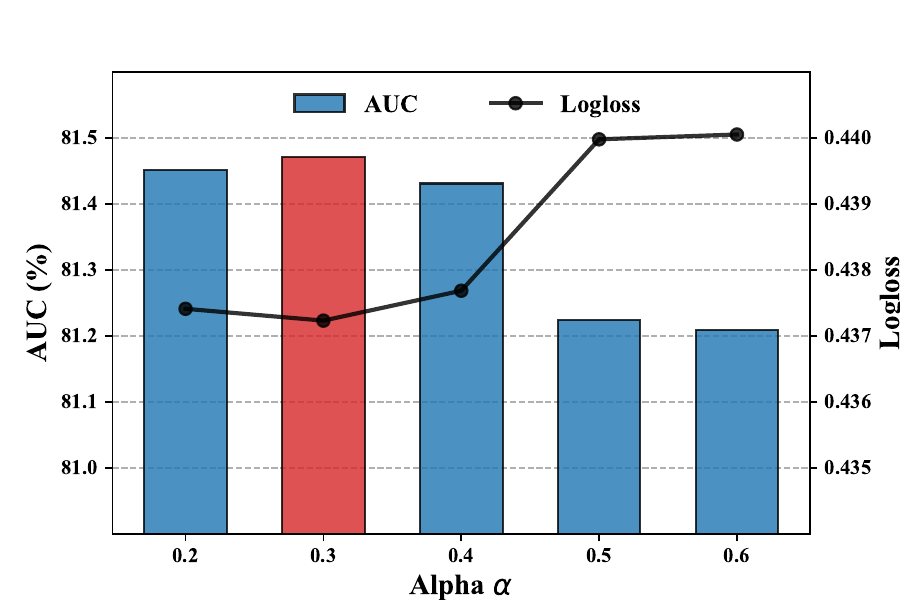}
        \centerline{(a) Criteo}
    \end{minipage}
    \begin{minipage}[t]{0.475\linewidth}
        \centering
        \includegraphics[width=\textwidth]{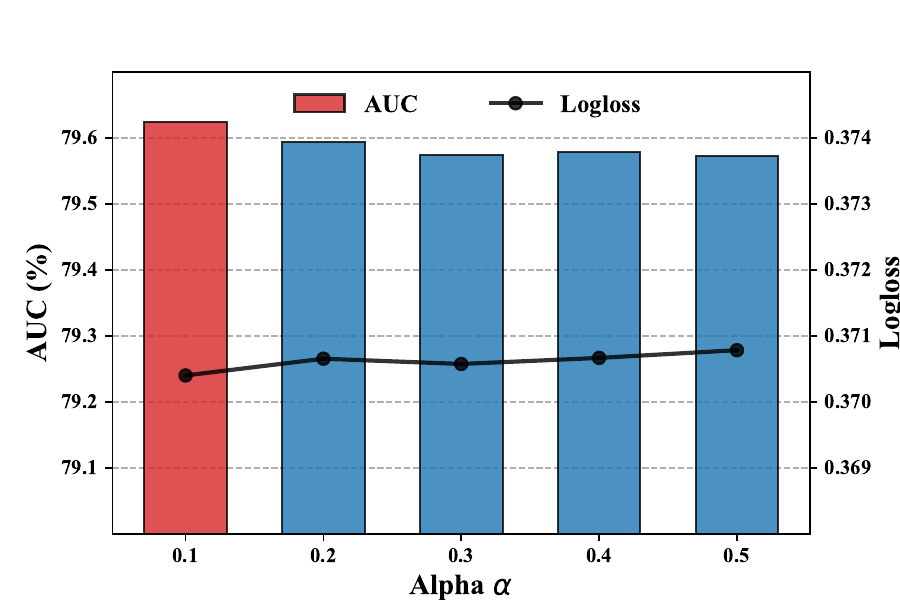}
        \centerline{(b) Avazu}
    \end{minipage}
    \begin{minipage}[t]{0.475\linewidth}
        \centering
        \includegraphics[width=\textwidth]{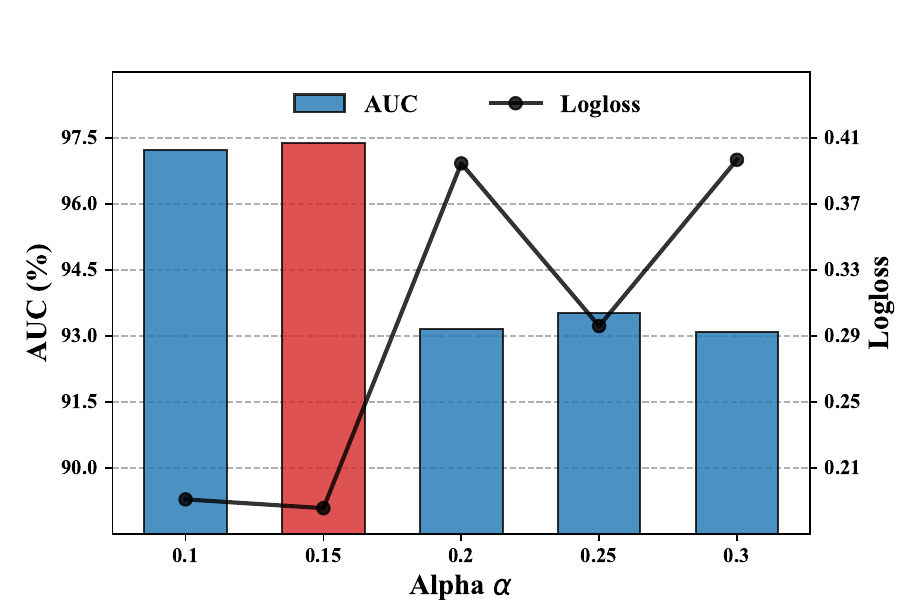}
        \centerline{(c) MovieLens}
    \end{minipage}
    \begin{minipage}[t]{0.475\linewidth}
        \centering
        \includegraphics[width=\textwidth]{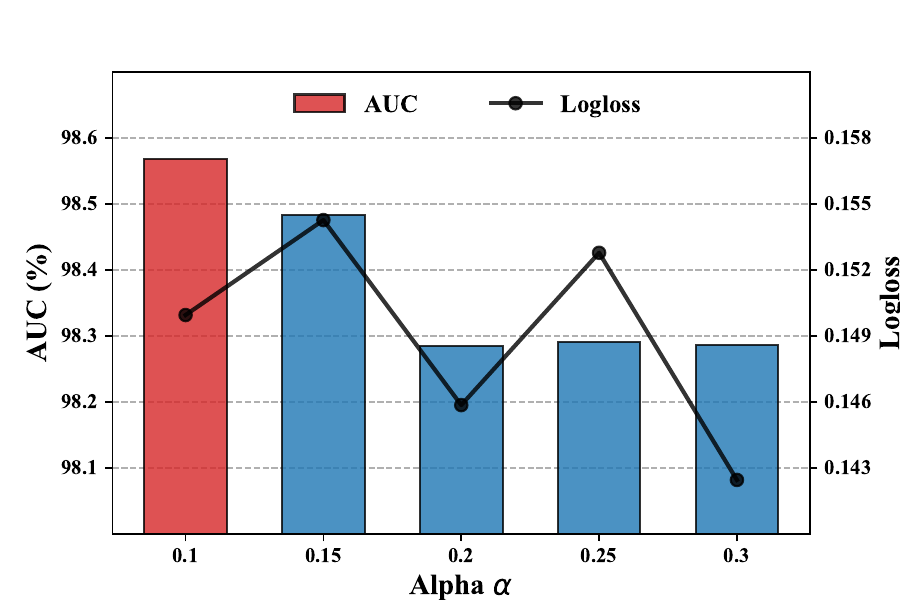}
        \centerline{(d) Frappe}
    \end{minipage}
    \captionsetup{justification=raggedright}
    \caption{Performance comparison of different weights of contrast loss on Criteo (a), Avazu (b), MovieLens (c), and Frappe (d) datasets.}
    \label{figure 7}
    \vspace{-0.5em}
\end{figure*}

\subsubsection{Impact of $\alpha$ in the $\mathcal{L}_{c l}$} We make modifications to the contrastive loss weight on the Avazu and Criteo datasets with a step size of 0.1, and in the smaller datasets MovieLens and Frappe, the modification step size is set to 0.05. The results are depicted in Figure \ref{figure 7}. On the Criteo dataset, it's observed that the model achieves better performance when the contrastive loss weight is set to 0.2 or 0.3. Subsequently, as the weight increases, the model's performance starts to decline. Notably, On the Movielens dataset, due to its limited three feature fields, the model is more sensitive to hyperparameter variations. Therefore, there is a substantial performance change when the weight is increased from 0.15 to 0.2. Compared to the MovieLens dataset, the performance of different values for the parameter $\alpha$ on the Avazu dataset is more stable. However, there is still a trend of declining performance with the increase in weight values. It's worth mentioning that compared to the simMHN model, when $\alpha$=0.1, the model's AUC performance improves by 0.48\%. This demonstrates the double-edged nature of diversity. On one hand, it can help the model capture additional information, and on the other hand, it can easily introduce noise signals. On the Frappe dataset, the model's performance peaks when $\alpha$=0.1, after which the model's performance shows a precipitous decline at 0.2, consistent with the trend observed on the MovieLens dataset. Looking at the overall Figure \ref{figure 7}, we find that when the CETN model achieves optimal performance in various datasets, the value of $\alpha$ is between 0.15 and 0.3. This suggests that although diversity of information is necessary, it still needs to be managed moderately. Otherwise, it will bring unnecessary noise signals to the model in various semantic spaces.

\begin{figure*}[t]
    \begin{minipage}[t]{0.475\linewidth}
        \centering
        \includegraphics[width=\textwidth]{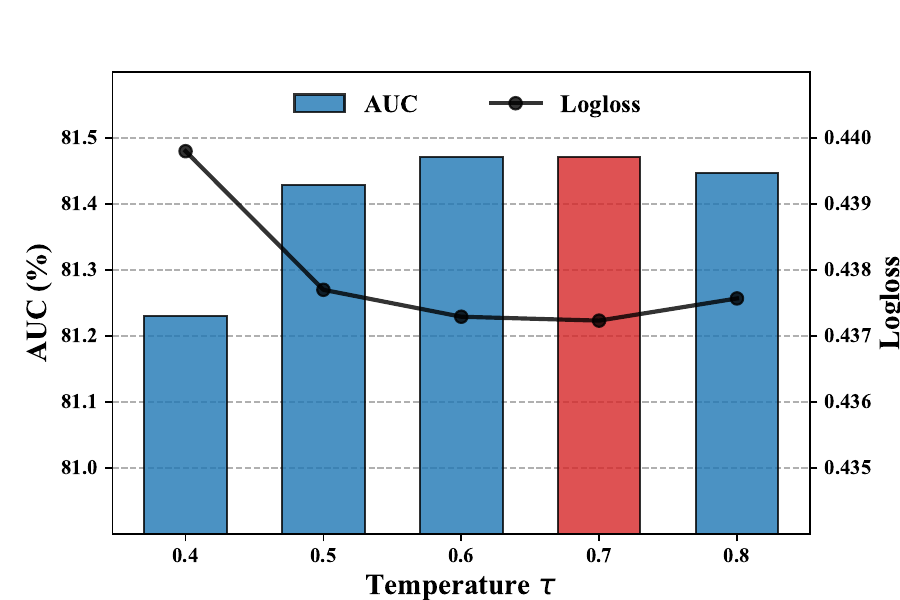}
        \centerline{(a) Criteo}
    \end{minipage}%
    \begin{minipage}[t]{0.475\linewidth}
        \centering
        \includegraphics[width=\textwidth]{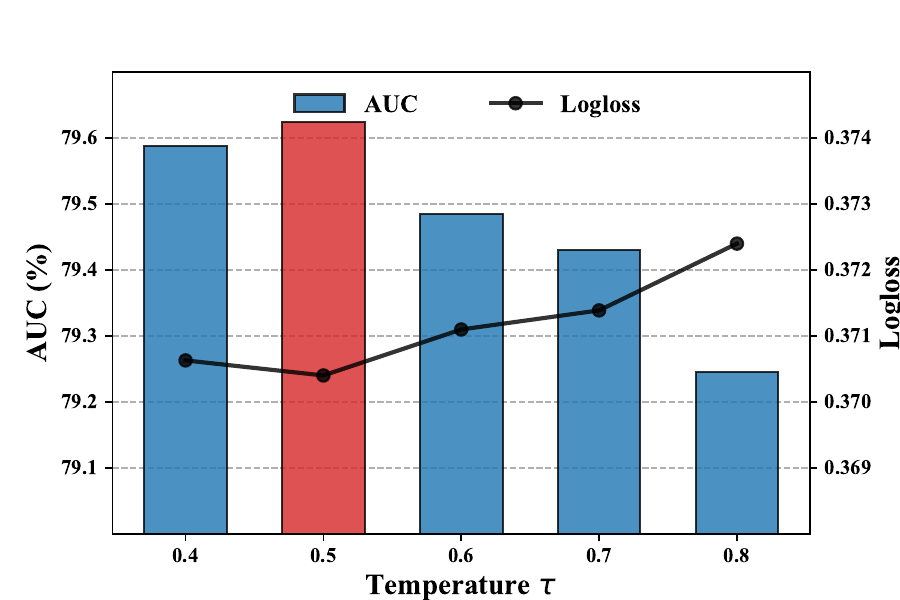}
        \centerline{(b) Avazu}
    \end{minipage}
    \begin{minipage}[t]{0.475\linewidth}
        \centering
        \includegraphics[width=\textwidth]{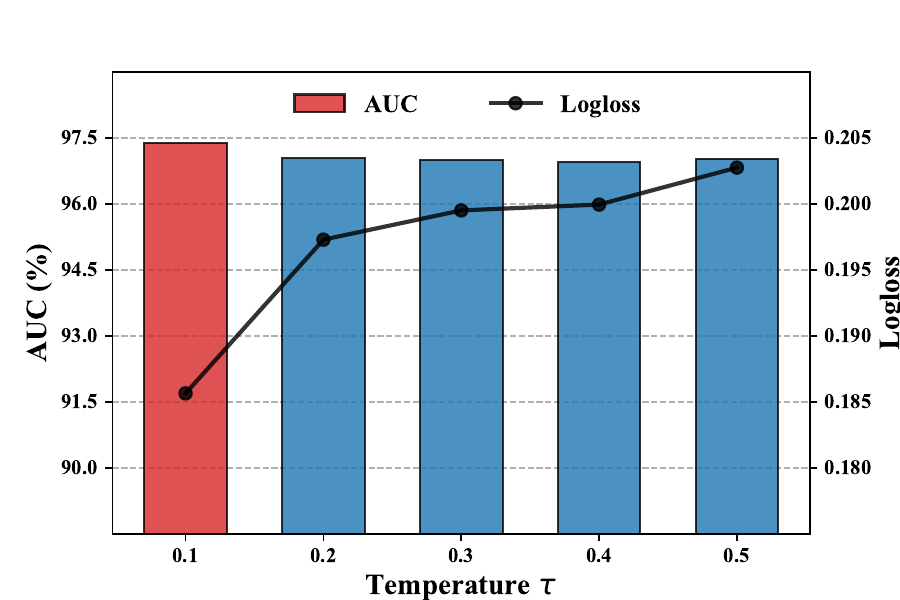}
        \centerline{(c) MovieLens}
    \end{minipage}
    \begin{minipage}[t]{0.475\linewidth}
        \centering
        \includegraphics[width=\textwidth]{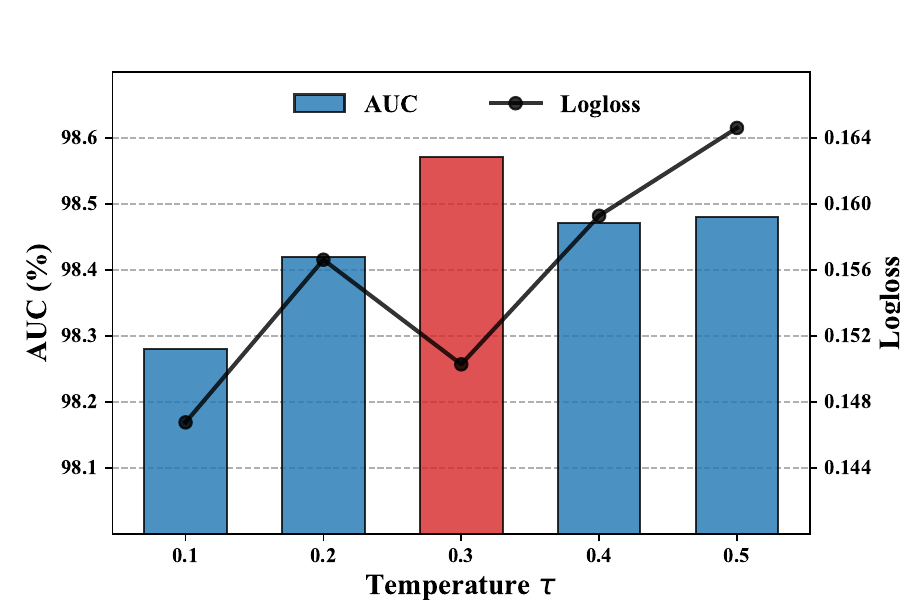}
        \centerline{(d) Frappe}
    \end{minipage}
    
    \captionsetup{justification=raggedright}
    \caption{Performance comparison of different temperature coefficients of CETN on Criteo (a), Avazu (b), MovieLens (c), and Frappe (d) datasets.}
    \label{figure 8}
\end{figure*}

\subsubsection{Impact of $\tau$} In many models based on InfoNCE loss \cite{simGCL, sgl, XsimGCL}, the temperature coefficient is commonly set to 0.2 by default. However, due to the sensitivity of click-through rate prediction tasks to performance metrics, we further explored its impact on the model's performance. We keep other parameters fixed and change the values for $\tau$ with a step size of 0.1 in the four datasets, as shown in Figure \ref{figure 8}. As can be observed, with an increase in the temperature coefficient, the model's performance gradually improves and then gradually decreases. However, MovieLens is an exception. To achieve good performance, the model requires an extreme temperature coefficient. This suggests that at this point, the model needs to rely on a stronger force to differentiate the semantic information between different input instances. In summary, we recommend fine-tuning the temperature coefficient of the contrastive loss within the range of 0.1 to 1.0 for optimal performance.

\subsection{Denominator-only InfoNCE vs InfoNCE}

\subsubsection{Role of Denominator-only InfoNCE} To make InfoNCE more suitable for feature interaction-based CTR tasks, we make modifications to it. By simplifying and deriving the formula for InfoNCE (Eq. \ref{6}), we can transform it into the following form:
\begin{align}
\label{23}
    &\sum_{i \in \mathcal{B}}-\log \frac{\exp \left(\text{sim}(\mathbf{V}_i^{\prime}, \mathbf{V}_i^{\prime \prime}) / \tau\right)}{\sum_{j \in \mathcal{B}} \exp \left(\text{sim}(\mathbf{V}_i^{\prime}, \mathbf{V}_j^{\prime \prime}) / \tau\right)}, \nonumber\\
   \Rightarrow&\sum_{i \in \mathcal{B}}-\log \frac{\exp \left(1 / \tau\right)}{\sum_{j \in \mathcal{B}} \exp \left(\text{sim}(\mathbf{V}_i^{\prime}, \mathbf{V}_j^{\prime \prime}) / \tau\right)}, \\
   \Rightarrow&\sum_{i \in \mathcal{B}}\log \left(\exp (\text{sim}(\mathbf{V}_i^{\prime}, \mathbf{V}_i^{\prime \prime}) / \tau) +\sum_{j \in \mathcal{B} /\{i\}} \exp \left((\text{sim}(\mathbf{V}_i^{\prime}, \mathbf{V}_j^{\prime \prime}) / \tau)\right)\right),
\label{24}
\end{align}
in Equation (\ref{23}), we discarded alignment, retaining only uniformity, thereby obtaining Denominator-only InfoNCE. In Equation (\ref{24}), we disregarded $1/\tau$, it becomes apparent that when the model optimizes this loss, it ensures that the model acquires dissimilar information across different semantic spaces, even if $\mathbf{V}_i$ and $\mathbf{V}_j$ originate from the same input. 

In some studies \cite{simGCL, XsimGCL, sgl}, this uniformity is interpreted as minimizing similarity. This perspective is especially prevalent in the context of graph neural networks, where the aim is to disperse nodes away from dense clusters in the representation space, leading to a more uniformly distributed representation. In fact, looking at this from the perspective of information capture, the reason for the improvement in model performance due to this uniform distribution can be simply attributed to the model capturing more diverse information. In other words, we ensure the diversity of the captured information through the InfoNCE loss function.

\subsubsection{Optimization Process.} 
\begin{figure}[]
    \begin{minipage}[t]{0.5\linewidth}
        \centering
        \includegraphics[width=\textwidth]{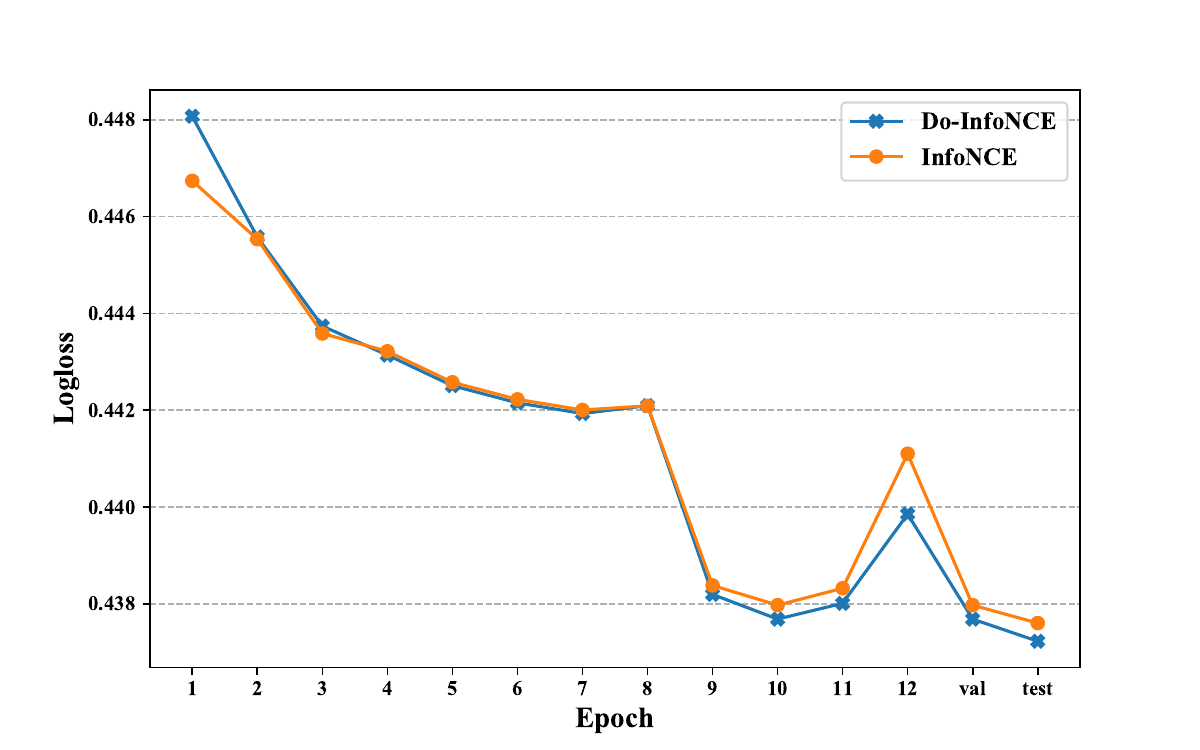}
        \centerline{(a) Criteo}
    \end{minipage}%
    \begin{minipage}[t]{0.5\linewidth}
        \centering
        \includegraphics[width=\textwidth]{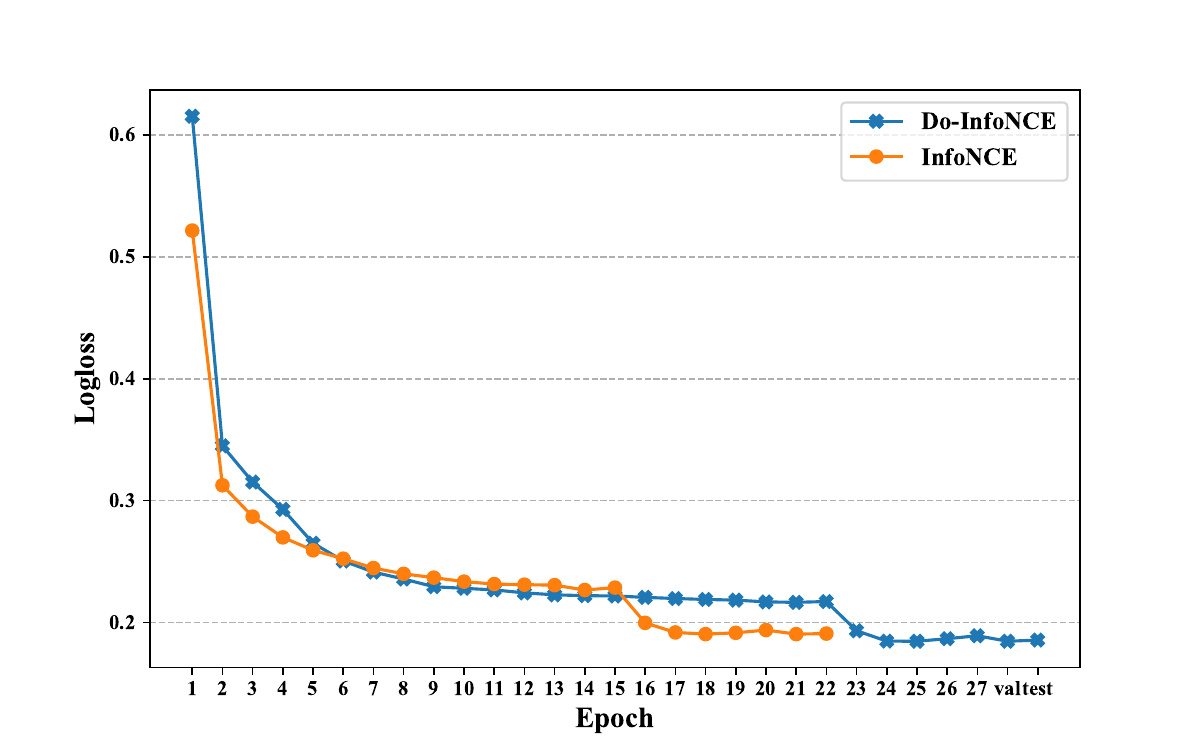}
        \centerline{(b) MovieLens}
    \end{minipage}
    \begin{minipage}[t]{0.5\linewidth}
        \centering
        \includegraphics[width=\textwidth]{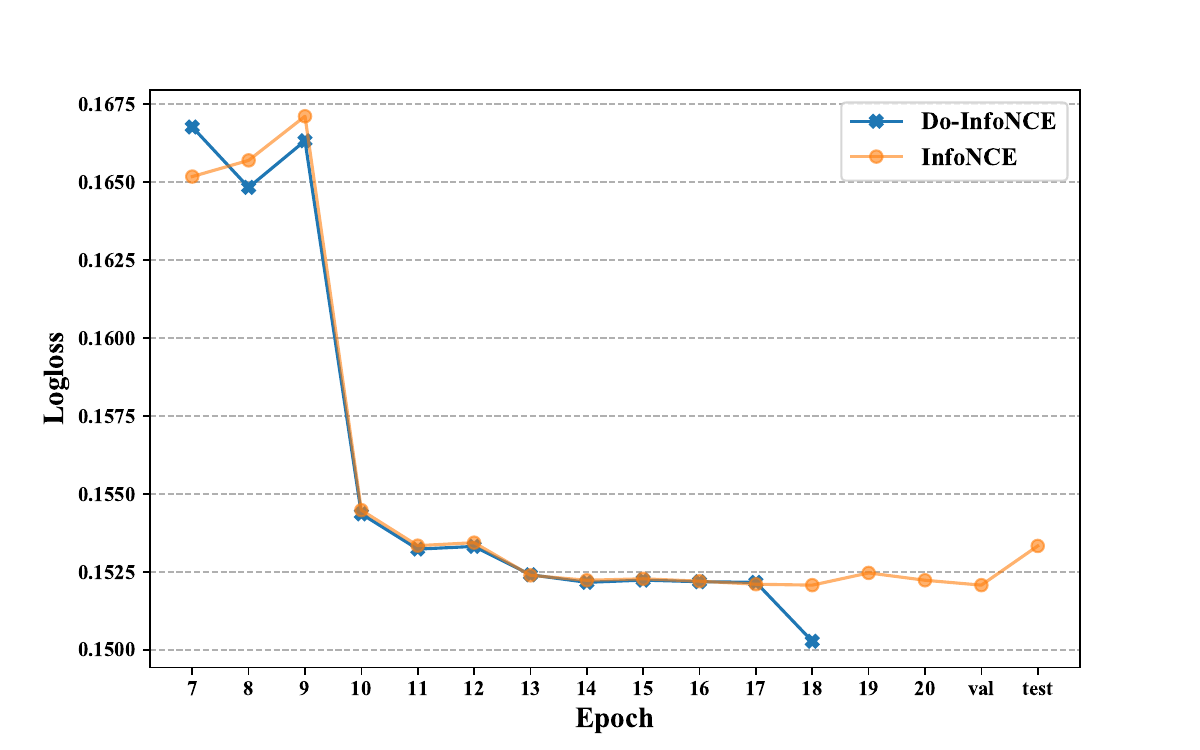}
        \centerline{(c) Frappe}
    \end{minipage}
    \captionsetup{justification=raggedright}
    \caption{Optimization Process of CETN on Criteo (a), MovieLens (b), and Frappe (c) datasets.}
    \label{figure 9}
\end{figure}

To further investigate the impact of Do-InfoNCE and InfoNCE on the model's loss optimization process, we visualize the model's training process, and the results are presented in Figure \ref{figure 9}. 
On the Criteo dataset, while InfoNCE initially yields better results in the first epoch, after lowering the learning rate with the Adam optimizer, the learning process of the model exhibits an issue which is insufficient capacity to capture information. Therefore, the optimization performance of InfoNCE is less favorable. On the Movielens dataset, with the early-stop patience set uniformly to two, InfoNCE even prematurely terminates the training process, so it obtains sub-optimal results. In the smaller Frappe dataset, Do-InfoNCE demonstrates superior performance. It stops training at the 16th epoch and achieves better results than InfoNCE in the test set (at the 18th epoch). In contrast, InfoNCE not only slows down the model's convergence but also leads to serious overfitting issues. Looking at the overall picture, our proposed Do-InfoNCE achieves better performance, which can aid the model in finding local optimum more effectively. 

\subsubsection{Visualization of information within semantic spaces.} 
\begin{figure*}[t] \scriptsize
    \begin{minipage}[t]{0.24\linewidth}
        \centering
        \includegraphics[width=\textwidth]{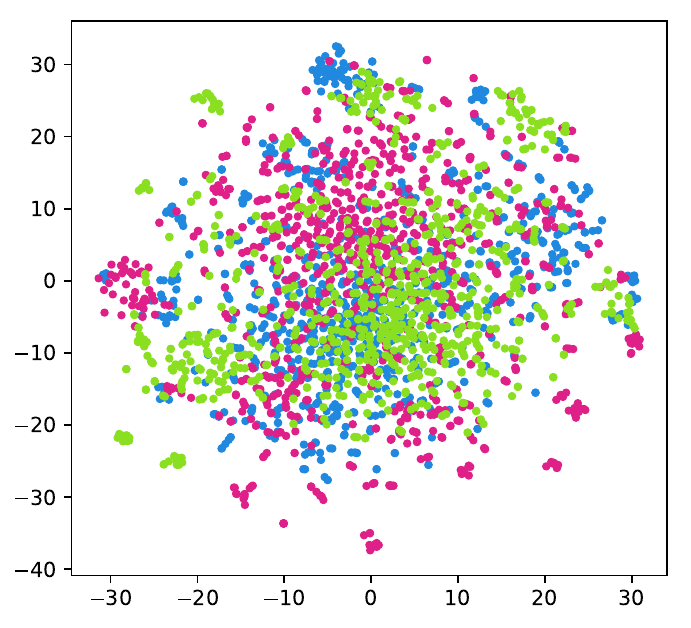}
        \centerline{(a) CETN on Frappe}
    \end{minipage}%
    \begin{minipage}[t]{0.24\linewidth}
        \centering
        \includegraphics[width=\textwidth]{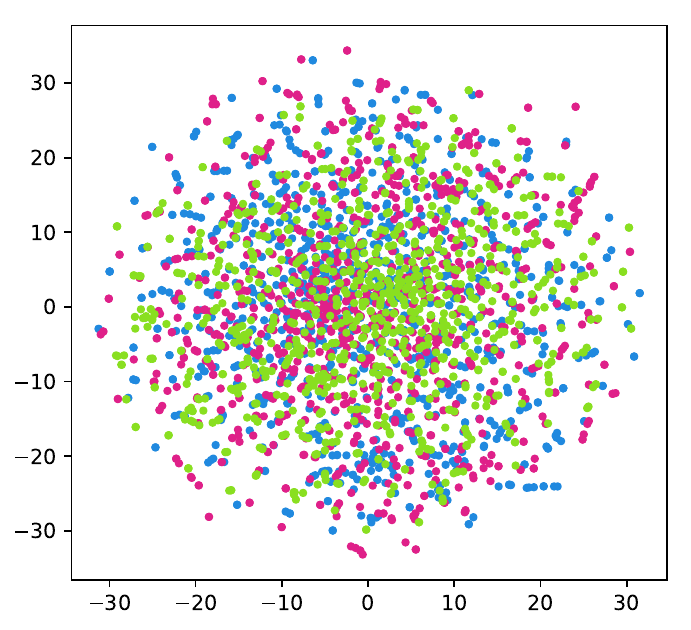}
        \centerline{(b) CETN on MovieLens}
    \end{minipage}%
    \begin{minipage}[t]{0.24\linewidth}
        \centering
        \includegraphics[width=\textwidth]{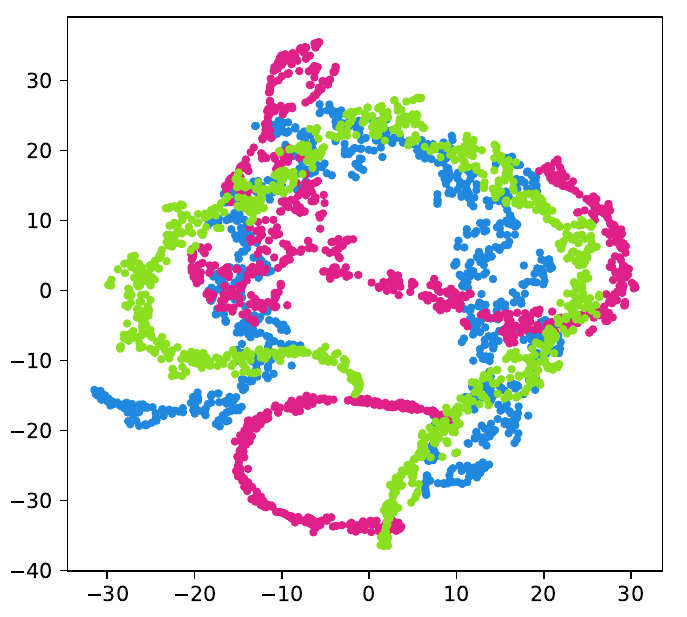}
        \centerline{(c) TriDNN on Frappe}
    \end{minipage}
    \begin{minipage}[t]{0.24\linewidth}
        \centering
        \includegraphics[width=\textwidth]{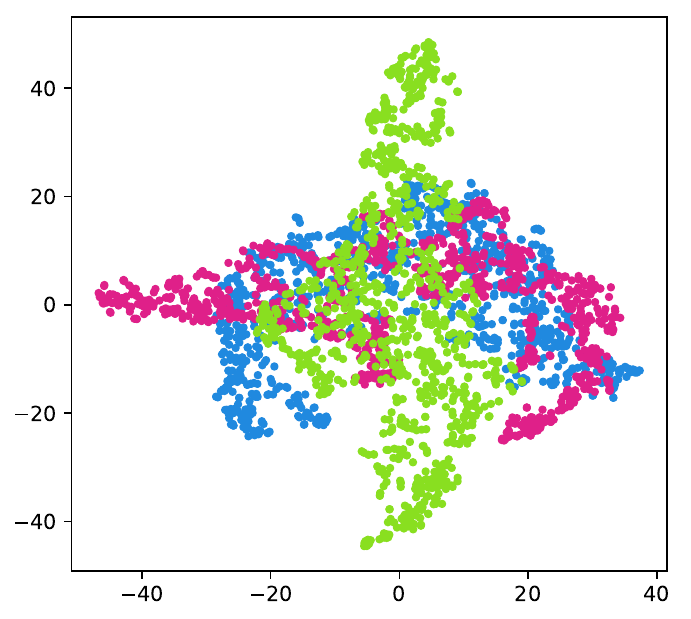}
        \centerline{(d) TriDNN on MovieLens}
    \end{minipage}
    \captionsetup{justification=raggedright}
    \caption{The visualization of information within semantic spaces on Frappe and MovieLens datasets. The three colors in the plot represent three semantic spaces. TriDNN denotes the use of three independent MLPs as subcomponents without employing self-supervisory signals.}
    \label{distribution}
\end{figure*}

To further explore the impact of self-supervisory signals on capturing feature interaction information within various semantic spaces, we randomly sampled 1,000 instances and visualized the information captured by the model, as shown in Figure \ref{distribution}. In Figure \ref{distribution} (a) and (b), the representation distribution is more dispersed due to the influence of $\mathcal{L}_{cl}$, indicating that the model has captured richer and more diverse information. On the other hand, in the case of TriDNN, the feature interaction information captured by the model across the three semantic spaces tends to be more similar. Consequently, in Figure \ref{distribution} (c) and (d), the representation distribution is more concentrated, suggesting that the model's acquired information is narrower, thus reducing its effectiveness.

\subsection{Through Network vs Residual Network}
\subsubsection{The Widespread Phenomenon of Shallow Networks in Recommender Systems.} In the field of computer vision, neural networks often tend to develop in a deeper direction \cite{resNet, googleNet}, but in the field of recommender systems, shallow networks are often sufficient to achieve the expected task objectives. For example, in graph neural network-based recommender systems \cite{lightgcn,simGCL,CVGA}, the number of layers in the graph neural network is often set to 3. Similarly, in the click-through rate prediction tasks based on feature interaction \cite{finalmlp, openbenchmark}, the number of layers in the MLP is often also set to 3. The primary reason researchers set it this way is due to the issues of over-smoothing and degradation. This leads to a rapid descent of the neural network as the number of layers increases, eventually causing the model to collapse. Moreover, this severe data sparsity issue can even lead to a unique phenomenon in the field of recommender systems known as the one-epoch phenomenon \cite{one-epoch}. Therefore, in CTR prediction tasks, neural networks often tend to widen rather than deepen.

\subsubsection{Reproduction of the Collapse Phenomenon in CTR prediction.} 
\begin{figure}[t]
    \begin{minipage}[t]{0.475\linewidth}
        \centering
        \includegraphics[width=\textwidth]{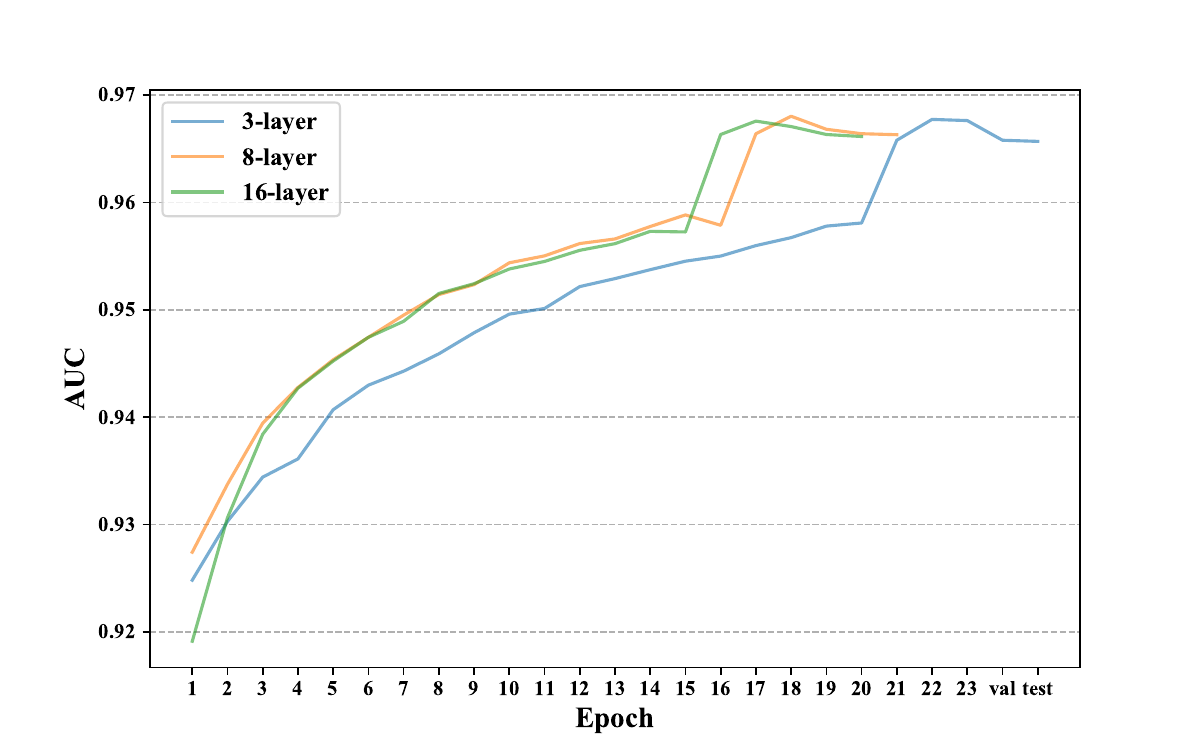}
    \end{minipage}%
    \begin{minipage}[t]{0.475\linewidth}
        \centering
        \includegraphics[width=\textwidth]{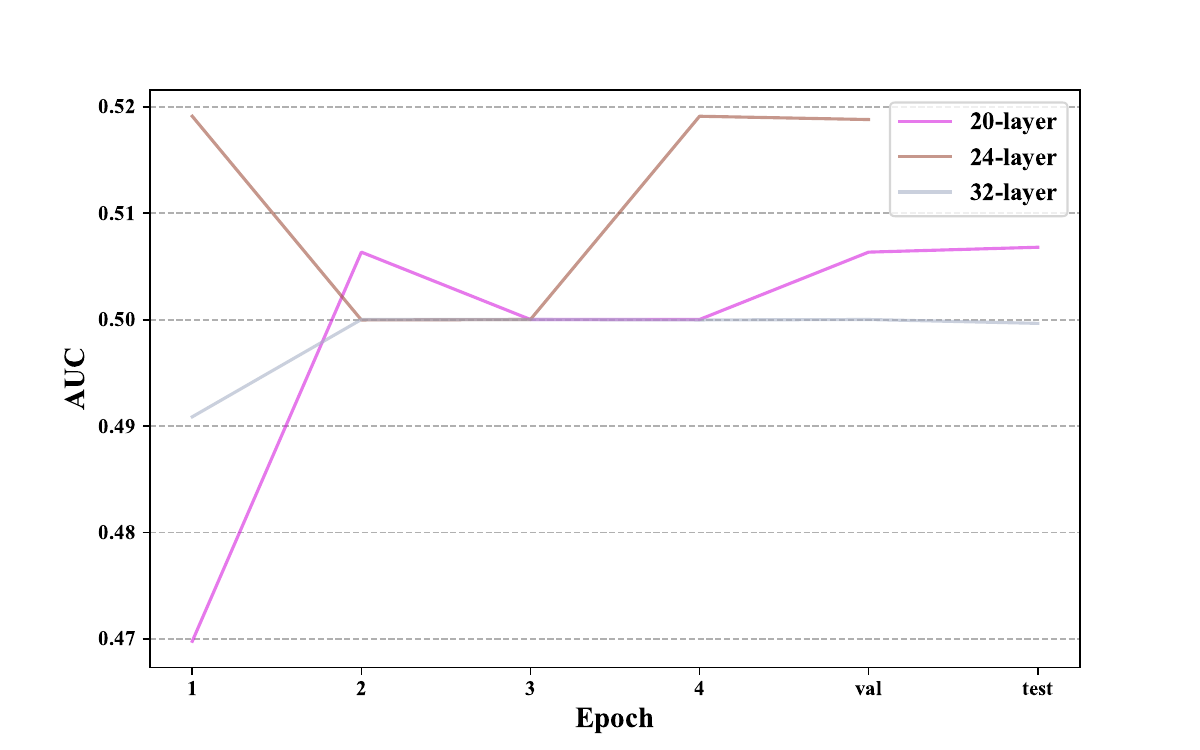}
    \end{minipage}
    \captionsetup{justification=raggedright}
    \caption{Optimization process of plain DNN with different number of layers on MovieLens datasets.}
    \label{figure 10}
\end{figure}

To verify the degradation phenomenon of deep neural networks in the field of recommender systems, we conducted experiments on the MovieLens dataset using MLP with different numbers of layers. We also visualized the training process of the model, as shown in Figure \ref{figure 10}. In order to intuitively observe the impact of the number of neural network layers on model training, we stopped using methods to prevent model overfitting, such as dropout and batch norm, and only retained $\textit{L}_{2}$ normalization to prevent the model from experiencing the one-epoch phenomenon. From the experimental results, it can be observed that as the number of MLP layers gradually increased from 3 to 8 and then to 16, the performance of the model does not show a significant improvement, but it does consume more computational resources. Therefore, further deepening the depth of the neural network is not a good choice. Next, when we increased the number of layers in the MLP to 20, the model experienced a collapse, and the AUC performance was close to 0.5. Even if we further deepened the depth of the MLP, the situation did not improve.

\begin{figure}[t]
    \begin{minipage}[t]{0.475\linewidth}
        \centering
        \includegraphics[width=\textwidth]{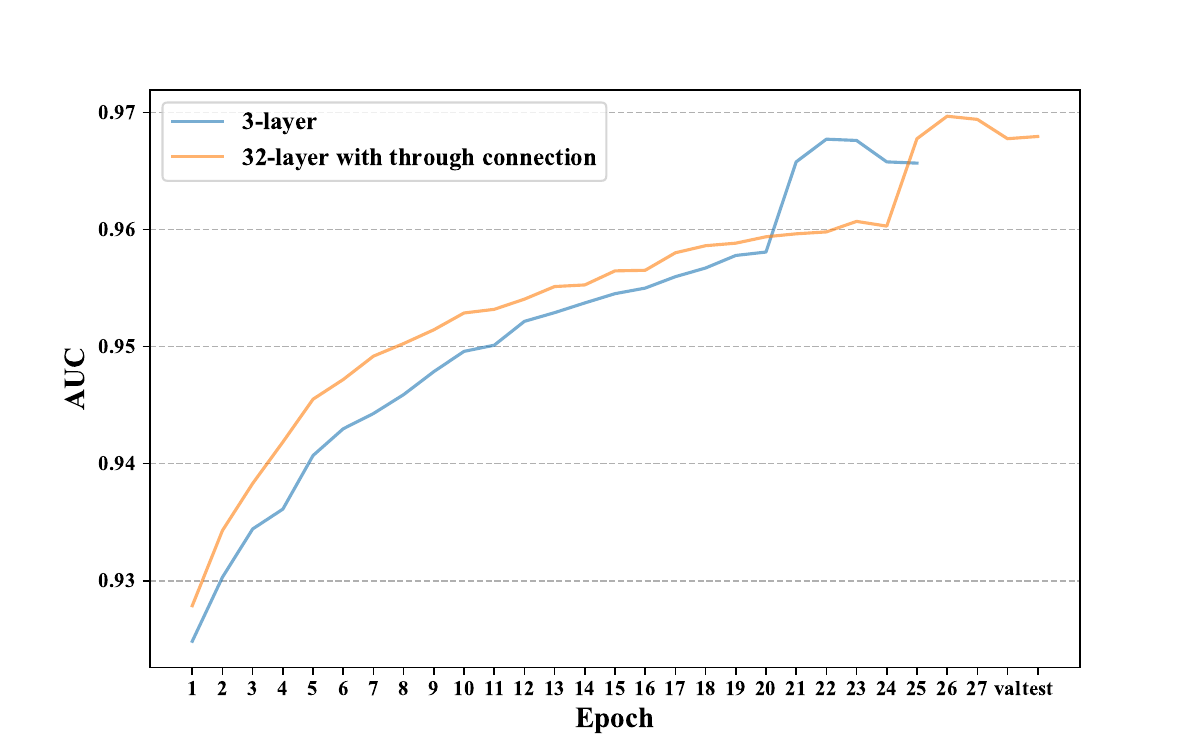}
    \end{minipage}%
    \begin{minipage}[t]{0.475\linewidth}
        \centering
        \includegraphics[width=\textwidth]{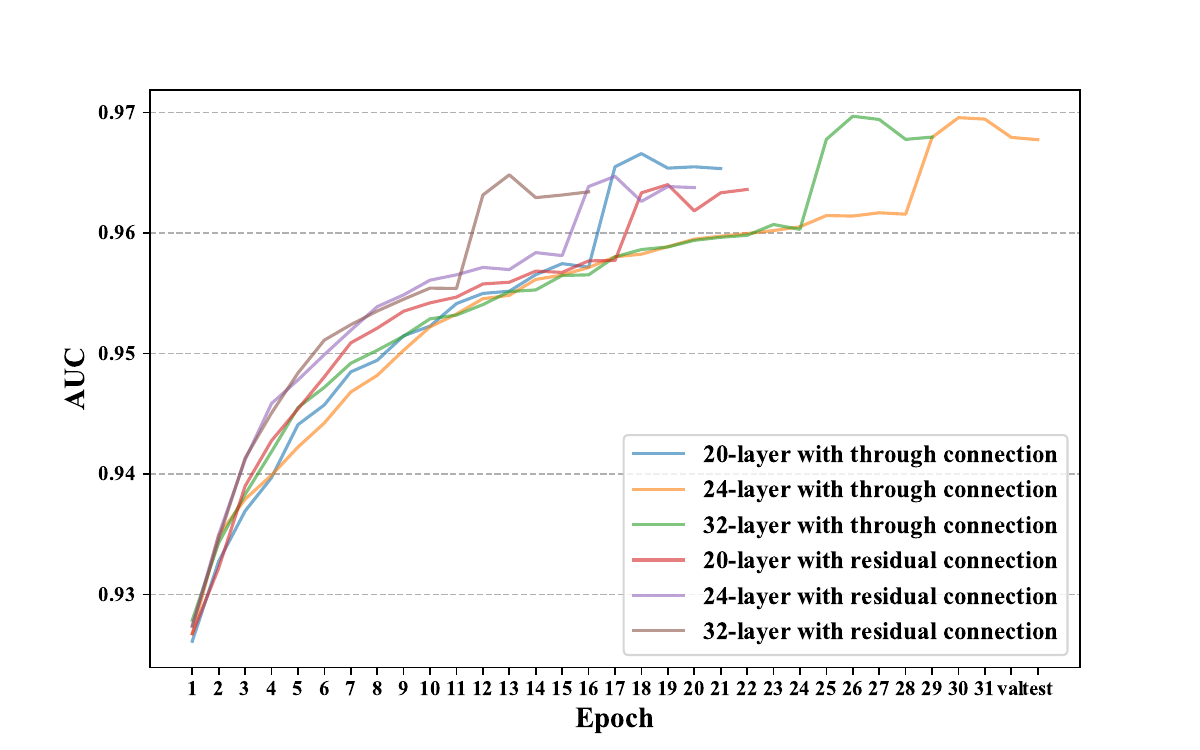}
    \end{minipage}
    \captionsetup{justification=raggedright}
    \caption{Optimization Process on MovieLens datasets. Left: comparison between a 3-layer plain DNN and a 32-layer DNN using through connection. Right: comparison between the residual network and the through network.}
    \label{figure 11}
\end{figure}

\subsubsection{Further Theoretical Analysis.} In previous research represented by residual networks, they use vertical, element-wise addition skip connections to pass information from earlier layers to later layers in the neural network, preventing the model from experiencing the degradation phenomenon. This residual connection can be simply defined as:
\begin{equation}  
\begin{aligned}
\mathbf{y}_{\text{deep}}&=\mathcal{F}\left(\mathbf{x}\right)+\mathbf{x},
\end{aligned}
\end{equation}
in this formula, $\mathbf{x}$ represents the input of a certain layer in the network, $\mathcal{F}$ is the transformation applied to $\mathbf{x}$ by the layer, and $\mathbf{y}$ is the output. The key idea here is that the output $\mathbf{y}$ is the sum of the input $\mathbf{x}$ and its residual (i.e., the difference between $\mathbf{x}$ and $\mathbf{y}$). Next, by horizontally generalizing it, we arrive at the fundamental definition of through connections:
\begin{equation}
\begin{aligned}    \mathbf{y}_{\text{shallow}}&=\mathcal{F}_{\text{shallow}}\left(\mathbf{x}\right), \\
\mathbf{y}_{\text{deep}}&=\mathcal{F}_{\text{deep}}\left(\mathbf{x}\right)+\mathbf{y}_{\text{shallow}},
\end{aligned}
\end{equation}
where we first obtain the output $\mathbf{y}_{\text{shallow}}$ through a shallow network, and then add the output of a deep network to the output of the shallow network element-wise to get $\mathbf{y}_{\text{deep}}$. In fact, the residual connections can be regarded as a special case of the through connections. We can further redefine the residual connection to the following form:
\begin{equation}
\begin{aligned}    \mathbf{y}_{\text{shallow}}&=\mathcal{F}_{\text{shallow}}\left(\mathbf{x}\right), \\
\mathbf{y}_{\text{deep}}&=\mathcal{F}_{\text{deep}}\left(\mathbf{\mathbf{y}_{\text{shallow}}}\right)+\mathbf{y}_{\text{shallow}},
\end{aligned}
\label{27}
\end{equation}
From Equation (\ref{27}), we can see that the difference between residual connections and through connections is simply that the input to $\mathcal{F}_{\text{deep}}$ has changed from raw $\mathbf{x}$ to $\mathbf{y}_{\text{shallow}}$.

\subsubsection{Validation of Effectiveness.} By changing the skip connections in the residual network from vertical to horizontal, we can generalize to the Through Network, which can help the model overcome the model collapse phenomenon brought about by the deepening of network layers. To demonstrate this hypothesis, we combined a 32-layer DNN  with a 3-layer DNN using through connections. The experimental results are shown in Figure \ref{figure 11} (left). What we can observe is that after using through connections, the 32-layer DNN not only avoids the model collapse phenomenon but also further improves the performance of the 3-layer DNN.

To compare the performance differences between through networks and residual networks in CTR prediction tasks, we conduct experiments using the layer settings where model collapse occurred. The results are shown in Figure \ref{figure 11}. One observation is that the performance of the model using through connections exceeds that of residual connections. The residual connections perform best at 24 layers, while through connections perform best at 32 layers. This implies that through connections can better utilize deep neural networks to fit the target function on sparse datasets. Another point worth mentioning is that in our subsequent experiments, as the depth of the network continued to increase, the performance of the DNN model using the through connections continued to improve. This suggests that the through connections not only prevent the model from collapsing in deeper settings but also contribute to better performance as depth increases.

\section{RELATED WORK}
\subsection{CTR Models Based on Feature Interactions}
\label{section 2}
Most existing CTR models based on feature interaction predominantly follow the Embedding \& Cross paradigm. They begin by encoding categorical and continuous features through embedding techniques. Subsequently, they employ a variety of complex cross-operations to augment the first-order feature data, achieving the goal of feature interaction, and ultimately enhancing model performance. MLP has played an indelible role in implicit feature interaction (Cross), greatly improving the benchmark performance of deep CTR models. However, many researchers have highlighted the inefficiency of MLP in learning product-based feature interactions (inner product, outer product, or Hadamard product)  \cite{mlpeffective,neuralvsmf}. 
Therefore, researchers have endeavored to employ additional feature data augmentation techniques to overcome the performance bottleneck of MLPs. Enhanced MLP models can primarily be categorized into two types: those enhanced based on stack structures and those enhanced based on parallel structures. 

NFM  \cite{NFM}, PNN  \cite{pnn1}, MaskNet  \cite{masknet}, and xCrossNet  \cite{Xcrossnet} employ a stack structure to enhance MLP-based CTR models. They aim to introduce explicit product-based feature interaction operations before using embeddings as inputs to the MLP, thereby breaking through the performance bottleneck of the MLP. From a semantic space segmentation perspective, this explicit product operation further enriches the feature interaction information within the current semantic space, resulting in improved performance. Wide\ \&\ Deep  \cite{widedeep}, DeepFM  \cite{deepfm}, DeepLight  \cite{deeplight}, FinalMLP  \cite{finalmlp}, xDeepFM  \cite{xdeepfm}, and DCN  \cite{dcn} employ a parallel structure. These models aim to introduce explicit feature interactions in a parallel manner to the simple MLP model. They achieve this by incorporating a fusion layer to capture both explicit and implicit feature interaction information simultaneously. While this parallel strategy of capturing feature interaction information in different semantic spaces has yielded promising results, it still fails to address the three issues we have identified, ultimately resulting in sub-optimal performance.

\subsection{Contrastive Learning for CTR Prediction}
To the best of our knowledge, until now, there has been very limited work combining contrastive learning with CTR prediction tasks based on feature interactions. This can occur because users often have a propensity for multiple interests in items, and it becomes challenging to definitively distinguish between positive and negative samples based solely on the user's click behavior. Therefore, traditional contrastive learning based on alignment and uniformity principles cannot be directly applied to CTR. With the rapid advancement of self-supervised learning in the fields of Natural Language Processing (NLP)  \cite{nlp1, nlp2} and Computer Vision (CV)  \cite{CV1, cv3, cv2}, due to the similarities between click-through rate prediction models based on user behavior sequences and NLP, they initially incorporate contrastive learning \cite{TOIS1,TOIS2,Survey4CL}. MISS  \cite{MISS} analyzes user behavior sequences and employs contrastive loss to enhance user interest representations at the feature level, thereby improving model performance. AQCL  \cite{AQCL} attempts to alleviate the problem of learning difficulties in representing the user's click history feature sequences in cold-start scenarios by introducing AQCL loss. CL4CTR  \cite{CL4CTR} introduced contrastive learning for the first time into feature interaction-based CTR prediction tasks. It proposes feature alignment and field uniformity for the feature field concept of CTR to enhance the quality of feature representation. However, it does not incorporate InfoNCE  \cite{InfoNCE} loss and fails to address the issues of diversity and homogeneity from an architectural perspective.

\section{CONCLUSION AND FUTURE WORK}
In this paper, we revisited the problem of effectively capturing feature interaction information from multiple semantic spaces, and composed the Simple Multi-Head Network (simMHN) with multiple Key-Value Blocks as parallel subcomponents. We then further enhanced simMHN around the two complementary principles of diversity and homogeneity, thereby proposing a new simple and effective CTR model, called the Contrast-enhanced Through Network (CETN). CETN builds upon simMHN by leveraging contrastive learning and through connections to further capture high-quality feature interaction information. Experimental results on four benchmark datasets validate the effectiveness of our proposed model. As we look toward future work, we are interested in refining the architecture of this model, seeking opportunities to make it even more simple and efficient.



\bibliographystyle{ACM-Reference-Format}
\bibliography{sample-base}










\end{document}